\documentclass[aps,pra,epsfigure,notitlepage,twocolumn,longbibliography,superscriptaddress]{revtex4-1}
\usepackage{svg}
\usepackage{bm} 
\usepackage{graphicx}
\usepackage{amsmath}    
\usepackage{latexsym}
\usepackage{amsfonts}   
\usepackage{amssymb}
\usepackage{comment}
\usepackage{array}      
\usepackage{epsfig}
\usepackage{txfonts}
\usepackage{xcolor}
\usepackage[colorlinks=true,linkcolor=blue,urlcolor=blue,citecolor=blue,pdfusetitle]{hyperref}
\usepackage{hyperref}
\usepackage{ulem}
\usepackage{tikz}
\usepackage{braket}

\usepackage[caption=false]{subfig}

\usepackage{appendix}

\begin{document}
\title{Minimal action shortcut to adiabaticity in a driven Kitaev chain: competing gaps in a topological transition at finite-time}

\author{Rafael Bentes de Sales}
\email{rb.sales@unesp.br}
\affiliation{Instituto de Física Teórica, Universidade Estadual Paulista,
CP 271, 01140-070 São Paulo, São Paulo, Brazil}

\author{Krissia Zawadzki}
\email{krissia@ifsc.usp.br}
\affiliation{Instituto de Física de São Carlos, Universidade de São Paulo,
CP 369, 13560-970 São Carlos, São Paulo, Brazil}

\begin{abstract}
    One of the main difficulties in preparing many-body ground states is achieving the target state through simple counterdiabatic controls. For critical systems crossing a transition to a topological phase, this task becomes even more challenging due to the closing of the gaps in multiple symmetry sectors. This is the case of the Kitaev chain, whose transition between the trivial and topological phases involves states belonging to different symmetry sectors.
    In this work, we apply the recently introduced minimal action shortcut to adiabaticity (MA-STA) to a Kitaev chain and propose a multi-step strategy to obtain the optimal control protocol to drive the system across its different phases. Our results show that high fidelities can be achieved through the adapted MA-STA at time scales much shorter than those of linear ramp protocols. We also compare the performance of both controls in suppressing work fluctuations. These findings may guide the design of STA protocols in many-body systems where competing energy scales and symmetries shape the global dynamics.
\end{abstract}
\maketitle
\section{Introduction}

In quantum systems, the analogue of the quasi-static limit is provided by an adiabatic dynamics \cite{ehrenfest1916adiabatische, Born, kato1950adiabatic}, in which the system remains in its instantaneous eigenstate throughout the evolution. Finite-time dynamics, however, induces transitions between instantaneous eigenstates, an effect that contributes to entropy production \cite{Landi_RMP,Alhambra_PRX}, also manifesting as enhanced work fluctuations \cite{Plastina_PRL,Zawadzki_2023}. These two effects are typically associated with decreased thermodynamic performance \cite{Campbell_2026}, as shown in the case of quantum heat engines operating non-adiabatic work strokes \cite{Denzler_2024, PhysRevB.101.054513,PhysRevE.105.044120}.

Shortcuts to adiabaticity (STA) provide a framework to address this problem by designing protocols that reproduce adiabatic dynamics in finite time \cite{TORRONTEGUI2013117, Gu_ry_Odelin_2019, STAcontrol_tutorial}. Among these, counter-diabatic (CD) \cite{Berry_2009, demirplak} driving offers, in principle, an exact solution by introducing additional control fields that enforce adiabatic evolution at arbitrary time scales \cite{PhysRevLett.118.100601}. Despite its generality, its implementation is generally impractical in many-body systems \cite{STAcontrol_tutorial}. It requires detailed knowledge of the full instantaneous spectrum and eigenstates of the Hamiltonian, which becomes prohibitive as the complexity of the spectrum grows rapidly with system size. Moreover, the resulting control fields are typically highly non-local and experimentally inaccessible in both analog and digital quantum simulation platforms. These difficulties are further exacerbated in systems undergoing quantum phase transitions \cite{PhysRevLett.109.115703, PhysRevB.72.161201,  PhysRevLett.119.060201, Polkovnikov_2011,PhysRevB.72.161201}, where energy gaps close and the adiabatic time scales diverge.

In recent years, considerable effort has been devoted to extending STA and CD techniques to many-body and critical systems \cite{del_Campo_2013, Sels_2017_CD, Claeys_PRL_2019, Ieva_PRX_localCD, STAcontrol_tutorial}. Approaches based on 
variational counterdiabatic (CD) driving \cite{Saberi_2014}, reinforcement learning \cite{Duncan_2025}, uniform adiabaticity \cite{Quan_2010}, dynamical invariants \cite{Espinos} and fast quasiadiabatic dynamics \cite{PhysRevA.92.043406, Martinez-Garaot:17}, and other approximate schemes \cite{STAcontrol_tutorial} have been proposed to mitigate the complexity of exact CD protocols. These methods have also been explored in the context of quantum thermodynamics \cite{Campbell_2026}, where adiabatic evolution translates into reduced excess work and improved performance \cite{beau2016scaling, Hartmann2020_PRResearch}. 

A recently proposed approach, known as the minimal action shortcut to adiabaticity (MA-STA) \cite{Kazhybekova_2022}, provides an alternative route to constructing efficient yet simple control protocols in many-body critical systems. This method reformulates the problem in terms of the minimization of an adiabatic action, an object that depends only on the relevant energy gaps of the system, leading to simpler control fields. The MA-STA has been successfully applied to paradigmatic models such as the Landau–Zener problem \cite{PhysRevA.65.042308} and the transverse-field Ising (TFIM) chain \cite{Kazhybekova_2022}.

While STA protocols have been extensively studied in systems characterized by a single dominant gap, their behavior in the presence of multiple competing gaps and topological transitions remains less explored. In this work, we extend the MA-STA framework to the Kitaev chain \cite{Kitaev_2001}, a system particularly suitable for investigating the interplay between many-body dynamics, topology, and finite-time control, as well as to test the MA-STA method in a dynamical setting characterized by multiple relevant energy scales associated with different momentum sectors. 
By analyzing protocols that drive the system across topological phase transitions, we show that the presence of competing gaps imposes nontrivial constraints on minimal-action protocols, requiring a multi-step strategy.
To assess the performance of the protocol, we consider both state fidelity and work statistics.

This paper is structured as follows.  In Sec. \ref{sec:ma-sta}, we present the MA-STA approach proposed in Ref. \cite{Kazhybekova_2022} and summarize previous results obtained from spin chains. In Sec. \ref{sec:kitaev_chain}, we present the Kitaev chain model, discussing its most relevant properties, and the strategies required to adapt the MA-STA protocol for an evolution in which the system undergoes a topological transition. Results for the fidelity and work statistics are discussed in Sec. \ref{sec:results}, and, finally, in Sec. \ref{sec:conclusions}, we draw our conclusions. Additional details are provided in the appendices \ref{sec:appendix_DELTA} and \ref{sec:appendix_A}, where we focus on the influence of the superconducting gap parameter on the protocol's effectiveness, and present a more detailed derivation of the figures of merit using the bulk approximation.

\section{Minimal action shortcut-to-adiabaticity}
\label{sec:ma-sta}
We are interested in the evolution of an initial state towards a target state through a time-dependent Hamiltonian $\hat{H}(g(t))$ controlled by a scalar drive $g(t)$.  Depending on the speed of this process, the system can experience transitions between different instantaneous energy eigenstates. According to the adiabatic theorem \cite{PhysRevA.80.012106}, the generation of unwanted excitations can be avoided by a sufficiently slow dynamics, defining a time scale that is inversely proportional to the minimal spectral gap. 

There are several scenarios in which these time scales render the evolution impractical  \cite{PhysRevLett.119.060201, PhysRevB.72.161201}. For systems exhibiting critical behavior, characterized by a spectral gap closure, the adiabatic condition breaks down. This is the case for many-body systems crossing a quantum phase transition at $g=g_c$, for which the time necessary to carry out an adiabatic evolution diverges.

In recent years, a number of shortcut to adiabaticity approaches to critical systems have been proposed to overcome these difficulties \cite{STAcontrol_tutorial}. Among them, a promising one is based on minimization of the adiabatic action \cite{Kazhybekova_2022}.  At the core of the MA-STA is the idea that the minimal gaps of the system set a timescale for reaching adiabaticity. Consequently, high-fidelity controls of the ground state can be achieved by varying a scalar parameter $g(t)$ while accounting for the timing and nature of the spectral gap closure. A key feature of this method is that it requires only the energy eigenvalues, rather than the full set of eigenvectors. This represents a crucial advantage in many-body systems, which often lack exact analytical solutions; in such cases, approximate numerical or analytical energy spectra are sufficient to achieve near-adiabatic dynamics from accessible information \cite{Campbell_2026}.

Reminiscent of \cite{Rezakhani_2009}, the adiabatic action is defined as
\begin{equation}
    \label{eq:action}
    S= \int_0^\tau dt\frac{\hbar^2||\partial_t \hat{H}(t)||^2}{\Gamma^4(t)},
\end{equation}
where $\tau$ is the total duration of the protocol, $||\cdot||$ the Frobenius norm and $\Gamma(t)$ the relevant energy gap, a quantity that serves as an estimate of the real energy gap that controls the dominant non-adiabatic transitions during the evolution.

The adiabatic action $S$ has units of time and its definition formally encodes the usual adiabatic conditions \cite{messiah1962quantum, PhysRevA.80.012106}: it increases for systems with smaller energy gaps $\Gamma(t)$ and larger system sizes, characterized by a larger Frobenius norm, and is proportional to the rate of change of the Hamiltonian, $\partial_t \hat{H}(t)$. While the time scales required to reach a target state with high fidelity are not explicitly defined, the adiabatic action suggests that the total evolution time can be optimized. Specifically, one can reduce the duration by maintaining minimal derivatives only at the critical instants $t_c$ where the gap $\Gamma(t_c)$ is small without entirely violating adiabaticity.

Therefore, the adiabatic action provides a new interpretation of the adiabatic condition for many-body critical systems and, at the same time, recasts the problem of finding the control function $g(t)$ into a minimization:
\begin{equation}
    \frac{\delta S}{\delta g} = 0.
\end{equation}
By solving the corresponding Euler-Lagrange equation, one finds the optimal protocol.

Note that the MA-STA provides a straightforward procedure for optimizing the control once the relevant gaps of the system are known. In principle, the method is suitable for treating any many-body system whose gaps can be estimated, being these system critical or not. A particular case in which the gaps can be obtained exactly includes systems described by analytically solvable (or models admitting an analytical solution). The TFIM studied in the original paper falls in this category. 

Here, our goal is to analyze the performance of the MA-STA to a exactly solvable model used to explore topological phase transitions: the Kitaev chain. We introduce the Hamiltonian and discuss its phases in the next section.

\section{Kitaev chain}
\label{sec:kitaev_chain}
The Kitaev chain \cite{Kitaev_2001, Leumer_2020} is an elementary model used to investigate topological superconductivity and Majorana zero modes (MZM) in nanowires. It can be mapped from a one-dimensional spin-$1/2$ system governed by Ising-like interactions in a transverse field with open boundary conditions (OBC). Via a Jordan-Wigner transformation, the transformed Hamiltonian describes spinless fermions with tight-binding interactions and a superconducting term. Explicitly,
\begin{equation}
\begin{split}
    \hat{H} = \sum_{j=1}^N \bigg[ -\mu(t) \hat{c}_j^\dagger \hat{c}_j -& \omega \left(\hat{c}_j ^\dagger \hat{c}_{j+1} + \hat{c}_{j+1}^\dagger \hat{c}_j \right) \\
    +& \left( \Delta^* \hat{c}_j^\dagger \hat{c}_{j+1}^\dagger + \Delta \hat{c}_{j+1} \hat{c}_j \right) \bigg],
\end{split}
\end{equation}
where $\hat{c}^\dagger_j/\hat{c}_j$ are creation/annihilation fermionic operators at site $j$, $\mu(t)$ is the chemical potential, which plays the role of the control parameter $g(t)$ -- hence the time dependence --, $\omega$ is the hopping amplitude and $\Delta$ is the pairing amplitude, associated with the superconducting gap. 

\begin{figure*}
    \centering
    \begin{minipage}[c]{0.49\textwidth}
        \centering
        \vspace{1cm}
        \subfloat[]{\includegraphics[width=\linewidth]{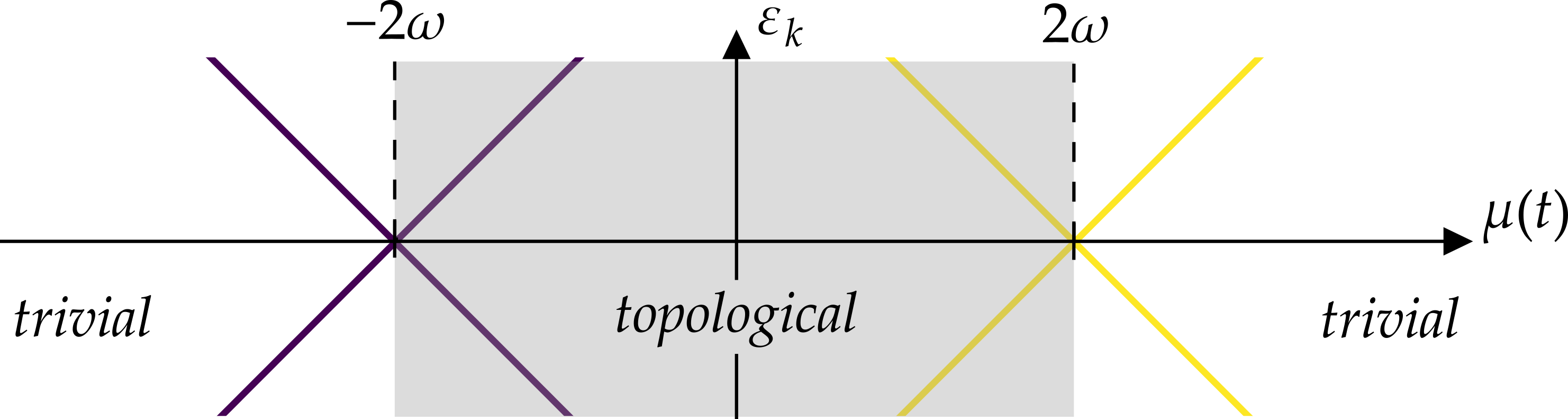}
        \label{fig:kitaev_phase_diagram}}
        \vspace{1.05cm}
        \subfloat[]{\includegraphics[width=\linewidth]{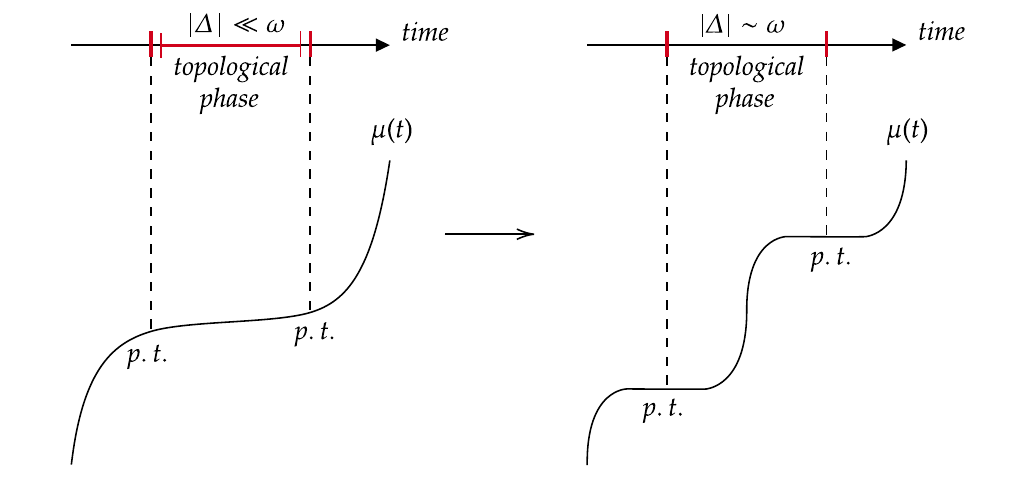}
        \label{fig:ma_sta_prot_drawing}}
    \end{minipage}
    \hfill
    \begin{minipage}[c]{0.49\textwidth}
        \centering
        \subfloat[]{\includegraphics[width=\linewidth]{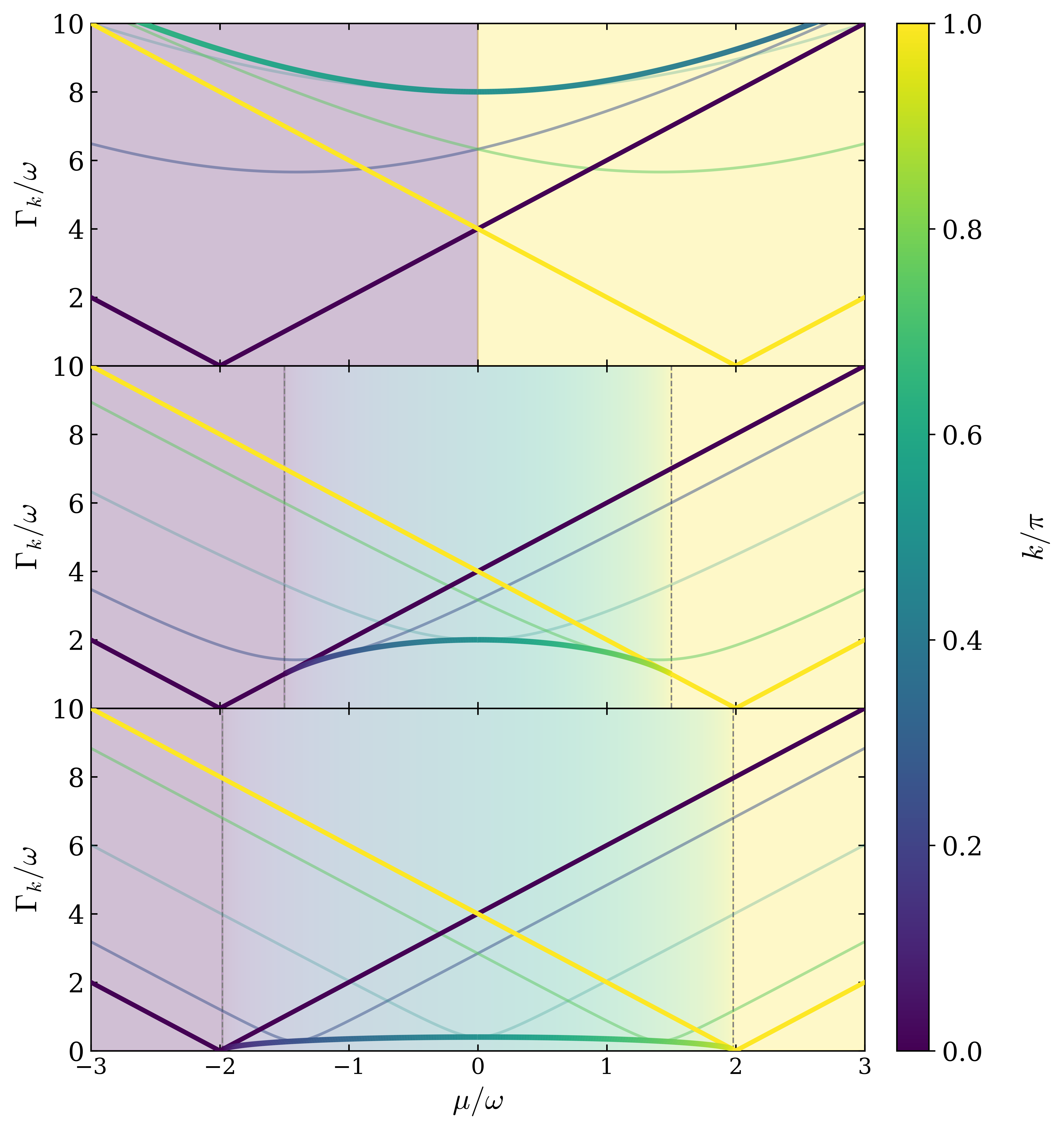}
        \label{fig:kitaev_levels_colored}}
    \end{minipage}

    \caption{(a) Evolution of the energy eigenvalues $\varepsilon_0^\pm$ (purple) and $\varepsilon_\pi^\pm$ (yellow) across the phase diagram. (b) Schematic of the minimal action protocols from $\mu_0 < -2\omega$ to $\mu_f > 2\omega$, crossing the entire topological phase. (c) Spectral gaps $\Gamma_k/\omega$ versus the chemical potential $\mu/\omega$ for $|\Delta| = 2\omega$ (top), $|\Delta| = 0.5\omega$ (middle), and $|\Delta| = 0.1\omega$ (bottom). In (a) and (c), the spectral gap closes at $\mu = -2\omega$ for the $k = 0$ subspace (purple solid lines) and at $\mu = 2\omega$ for the $k = \pi$ subspace (yellow solid lines), defining the boundaries between the trivial and topological phases of the Kitaev chain. In (c), the curve with varying color represents $\Gamma_{S_3}$, an energy gap solution that exists only when Eq. (\ref{eq:range_validity_ks3}) is satisfied (range delimited by vertical dashed grey lines). Within this regime, $\Gamma_{S_3}$ becomes the global minimum of the system if $|\Delta| < \omega$. Shaded background regions highlight the momentum sector associated with the smallest gap at each point. Note that the momentum of this global minimum gap changes discretely for strong pairing (top) but smoothly for weak pairing (middle and bottom). This gap structure dictates the protocol shapes shown in (b): for small $|\Delta|$ (left), the rate of change must remain low not only at the phase transitions (p.t.) but also throughout the extended interval given by Eq. (\ref{eq:range_validity_ks3}), in which $\Gamma_{S_3}$ is a minimum (highlighted in red). Conversely, for $|\Delta| \geq \omega$ (right), the relevant minimal gaps are localized solely at the phase boundaries, yielding a symmetric protocol with two distinct plateaus that allows for faster overall dynamics.}
    \label{fig:two_column_showcase_panel}
\end{figure*}

Under periodic boundary conditions (PBC), or in the infinite-wire limit, translational invariance allows us to define a good momentum quantum number $k$. Transforming to momentum space decouples the system into a single-particle problem defined by a collection of non-interacting momentum subspaces. This factorization reduces the full $2^N \times 2^N$ many-body Hamiltonian to a set of independent $2 \times 2$ matrices, significantly simplifying the computational complexity. Assuming an even number of sites $N$ and appropriate boundary conditions, the quantized momenta are given by $k=(2n-1)\pi/N$ for $n=1,2,\dots, N/2$. The resulting bulk excitation spectrum \cite{Leumer_2020},

\begin{equation}
\varepsilon_k^\pm = \pm \sqrt{\left( \mu(t) + 2\omega\cos{k} \right) ^2 + 4|\Delta|^2 \sin^2{k}},
\end{equation}
provides an exact description in the thermodynamic limit ($N \to \infty$). More importantly for our purposes, it serves as a highly effective approximation for the dynamics of large finite chains with open boundary conditions, successfully capturing the essential gap-closing features that drive the topological phase transitions.

This system has two topological phase transitions at $\mu(t)=\pm2\omega$, separating a topological phase, for $|\mu|<2\omega$, from the trivial phases, for $|\mu|>2\omega$ \cite{Alicea_2012}. The phase diagram is illustrated in Fig. \ref{fig:kitaev_phase_diagram}.

    

The existence of different phases can be explicitly identified on the Majorana representation, given by the operators
\begin{equation}
    \hat{a}_j+i\hat{b}_j \equiv \hat{c}_j^\dagger,
\end{equation}
in which each fermionic site $j$ is formed by two Majorana modes $a_j$ and $b_j$ -- quasi-particles that pair up to form the real particles, but might emerge at the ends of the chain on the topological phase. 

A non-local quantity that allows to distinguish topological and trivial phases is the String Order Parameter (SOP), defined as \cite{Chitov_2018}:
\begin{align}
    \hat{O}_x(m) &= \prod_{l=1}^{m-1}\left[i\hat{b}_l\hat{a}_{l+1}\right],\\
    \hat{O}_y(m) &= \prod_{l=1}^{m-1}\left[-i\hat{a}_l\hat{b}_{l+1}\right],\\
    \mathcal{O}^2_{x/y}&=\lim_{(n-m)\to \infty}|\langle \hat{O}_{x/y}(m) \hat{O}_{x/y}(n) \rangle|,
\end{align}
where $x$ or $y$ depends on whether $\Delta/\omega$ is respectively positive or negative.

\begin{figure}
    \centering
    \includegraphics[width=\linewidth]{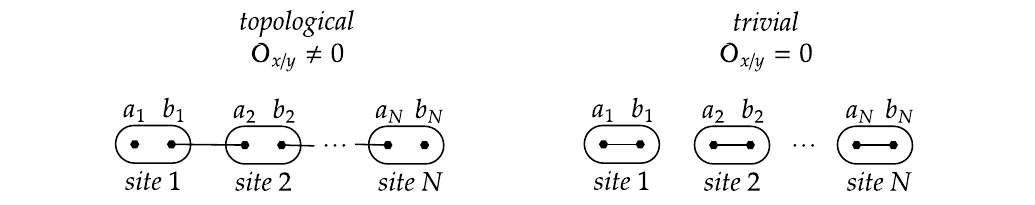}
    
    \caption{Kitaev Chain's phases. The main feature of the topological phase is the offset in the pairing of Majorana modes, which leaves MZMs unpaired at the ends of the chain. No unpaired quasi-particle is left in the trivial phase.}
    \label{fig:kitaev_phases}
\end{figure}

At the trivial phase $\mathcal{O}_{x/y}=0$ and the Majorana modes of the same fermionic site pair up to form $N$ fermions. In contrast, at the topological phase the SOP $\mathcal{O}_{x/y}\neq0$ and the pairing is offset by one -- modes of different fermionic sites pair up --, forming $N-1$ fermions and leaving unpaired modes at the ends of the chain (Fig. \ref{fig:kitaev_phases}). These edge modes, the so-called Majorana Zero Modes (MZMs) \cite{Alicea_2012}, combine non-locally to form the remaining $N$'th site and still allow two possible configurations on it, but with zero energy -- thus creating a degeneracy on the energy levels of the system, including the ground state.

The Kitaev Chain's Hamiltonian conserves parity, so it is split into two isolated sectors with even and odd parities, which are mutually inaccessible through non perturbative dynamics. In the trivial phase, where there is no degeneracy, the ground state has definite even parity. However, in the topological phase, the ground state is degenerate with different parities, and therefore does not have a well-defined parity.

We are interested in investigating an evolution in which the system evolves from the trivial to the topological phase or vice-versa. It is important to note that the critical point separates a regime with degeneracy from another where degeneracy is lifted, and these regimes are defined in distinct symmetry sectors. In order to apply the MA-STA it is crucial to keep track of all states involved in the transition: one needs to know which state on the trivial phase is connected to which state on the topological phase by perfectly adiabatic dynamics. For example, starting in the topological phase, a perfectly adiabatic phase transition to the trivial phase will take the even topological ground state to the least-energetic state of the even parity sector, which is the trivial ground state. On the other hand, the exact same dynamics will take the odd topological ground state to the least-energetic state of the odd sector, which is the first excited state.

In the following, we discuss which states should be considered in order to apply the MA-STA to the driven Kitaev chain.

\subsection{Relevant gaps and momenta subspaces}

A key point in the MA-STA is to avoid the smallest spectral gap over the entire spectrum. The solution of the Kitaev chain allows us to obtain the gaps in each momentum sector, which read
\begin{equation}
    \Gamma_k =\varepsilon_k^+ - \varepsilon_k^- = 2\sqrt{\left( \mu(t) + 2\omega\cos{k} \right) ^2 + 4|\Delta|^2 \sin^2{k}}.
\end{equation}

Our goal is to identify the momentum sector $k_m$ in which the smallest $\Gamma_k$ is found. To do that, we can consider the thermodynamical limit and minimize $\Gamma_k$ over continua $k$, and, then, find in the discretized case the $k_m$ which is closest to the minimum. Carrying out the minimization, we have:
\begin{equation}
    \left. \frac{\partial \Gamma_k}{\partial k} \right|_{k=k_m}=-\frac{4\sin k_m \left(\omega\mu(t)+2(\omega^2-|\Delta|^2) \cos k_m\right)}{\sqrt{4|\Delta|^2 \sin^2k_m + (\mu(t) + 2\omega \cos k_m)^2}}=0.
\end{equation}

There are three possible solutions: $k_{S_1}=0$ and $k_{S_2}=\pi$, which refer to the usual minimum and maximum values of the spectral gaps, and $k_{S_3}=\cos^{-1}\left( \omega\mu(t)/[2(|\Delta|^2-\omega^2)] \right)$.

The third solution is associated with a gap with a crucial role for the situation in which the system evolves in time, as it can yield an instantaneous value below those for $k=0$ and $k = \pi$. Notice it only exists for $|\Delta| \neq \omega$ on the given range:
\begin{equation}
    |\mu(t)| \leq 2(|\Delta|^2-\omega^2)/\omega.
    \label{eq:range_validity_ks3}
\end{equation}
For $|\Delta|>\omega$, the smallest gap is $\Gamma_0$ for $\mu<0$ and $\Gamma_\pi$ for $\mu>0$ -- for $\mu=0$, they are identical. On the other hand, for $|\Delta|<\omega$, the smallest gap's momentum value changes constinuously from $k=0$ to $k=\pi$ through the curve given by the third solution, $k_{S_3}=\cos^{-1}\left( \omega\mu(t)/[2(|\Delta|^2-\omega^2)] \right)$.

The instantaneous change in the minimal gap's momentum sector has not been addressed in previous works exploring MA-STA. Our goal here is to study the interplay between these minima. To this aim, we start by exploring the dependence of all three gaps with model parameters. Their analytical expressions are below:
\begin{align}
&\Gamma_{S_1}=2\left( \mu(t) + 2\omega \right), \\
&\Gamma_{S_2}=2\left( \mu(t) - 2\omega \right), \\
&\Gamma_{S_3}=4|\Delta|\sqrt{1-\mu^2(t)/(4\omega^2-4|\Delta|^2)}.
\label{eq:gaps_solutions_123}
\end{align}


    

Fig. \ref{fig:kitaev_levels_colored} displays full gap spectra with the model parameters and highlights the three gaps. We observe that the gap $\Gamma_{S_3}$ remains finite, but  if $|\Delta|$ is sufficiently small, $\Gamma_{S_3}$ also is -- Eq. (\ref{eq:gaps_solutions_123}) shows it is on the order of $|\Delta|$. Therefore, $\Gamma_{S_3}$ dominates over $\Gamma_{S_1}$ and $\Gamma_{S_2}$ and becomes the relevant gap for every value of $\mu(t)$ within this range.

An important consideration about the protocols driving the Kitaev chain across the topological and trivial phases of $\mu(t)$ refers to the feasibility of the MA-STA optimization. Since the protocol needs a low derivative on values at which there is a relevant gap, for small $|\Delta|$ it has to satisfy:
\begin{flalign}
    &&\frac{d\mu_{STA}(t)}{dt} \approx 0, &&&|\mu_{STA}(t)| \in\left\{ 2\omega, \left[0, 2(|\Delta|^2-\omega^2)/\omega \right] \right\},
\end{flalign}
which is a continuous interval (region between the dashed grey lines on the lower graph of Fig. \ref{fig:kitaev_levels_colored} and on the left of Fig. \ref{fig:ma_sta_prot_drawing}), thus making the usage of the MA-STA protocol $\mu_{STA}(t)$ impractible, as we would need a very large total time evolution $\tau$. 

Therefore, to get rid of the continuous relevant interval, it is necessary to tune the superconducting gap $|\Delta|$ such that the third solution $k_{S_3}=\cos^{-1}\left( \omega\mu(t)/[2(|\Delta|^2-\omega^2)] \right)$ no longer exists or corresponds to maximum gap. By fixing $|\Delta| \geq \omega$, the protocol can be applied two successive times to avoid only the $k=0$ and $k=\pi$ gaps, resulting on a symmetric total protocol with two plateaus, as seen on Fig. \ref{fig:ma_sta_prot_drawing}.


\subsection{Protocols for minimizing the action}

Following the prescription of \cite{Kazhybekova_2022}, to apply the minimization of the action in Eq. (\ref{eq:action}) for models mapped into two-level systems defined in momenta subspaces, we need the norm of the driven Hamiltonian as well as the relevant gaps in each subspace. For the Kitaev chain, the general action has the same form of that obtained for the transverse field Ising model, which reads
\begin{equation}
    S=\int_0^\tau dt \left[ \frac{\alpha \dot{g} \gamma^2}{(g-\beta)^2+\gamma^2} \right]^2,
    \label{eq:generalization_adaction}
\end{equation}
where the parameters $\beta$ and $\gamma$ refer, respectively, to the point of avoided crossing and the relevant gap $\Gamma_k$. The coefficient $\alpha$ is a constant scaling factor. Their explicit expressions depend on the model considered, and are specified below for the Kitaev chain.

Conversely, the optimal solution of the Euler-Lagrange equations obeying the constraints $G(0)=g_0$ and $G(\tau)=g_f$ is 
\begin{equation}
    G(s)=\beta + \gamma \tan \left[ (1-s)\tan ^{-1}\left( \frac{g_0 - \beta}{\gamma} \right) +s \tan ^{-1}\left( \frac{g_f - \beta}{\gamma} \right) \right],
\end{equation}
with re-escaled time $s=t/\tau$. 

Once the relevant gaps and the norm of the time derivative of the driven Hamiltonian are identified, we can carry out the minimization of the action for the Kitaev chain. The latter have been discussed in the previous section, while the first yields 
$||\partial_t \hat{H}||= \sqrt{2} \dot{\mu}(t)$. Therefore, the adiabatic action for the Kitaev chain takes the form 
\begin{equation}
    S=\int_0^\tau dt \frac{\dot{\mu}^2(t)}{8\left( (\mu + 2\omega \cos k)^2 + 4 |\Delta|^2 \sin^2k\right)^2},
\end{equation}
where we identify $\alpha = (32 |\Delta|^2 \sin^2(\pi/N) )^{-1}$, $\beta = -2 \omega \cos k$ and $ \gamma = 2 |\Delta| \sin k$.

We have to be cautious about the initial and final conditions: if a symmetric protocol starts and ends at the trivial phase, with $\mu(0)=\mu_0<-2\omega$ and $\mu(\tau)=-\mu_0>2\omega$, two phase transitions are crossed. In that case, the optimal protocol for varying the chemical potential in a way to minimize the action is split into two:
\begin{align}
    \mu(t) &= \left\{ \begin{array}{ll}
         \beta_- + \gamma_- \tan\begin{bmatrix} \left(1-\frac{t}{\tau/2}\right)\tan ^{-1}\left( {(\mu_0 - \beta_-)}/{\gamma_-} \right)\\
         \hfill +\frac{t}{\tau/2} \tan ^{-1}\left( {- \beta_-}/{\gamma_-} \right) \end{bmatrix}, & t\leq \tau/2, \\
         \beta_+ + \gamma_+ \tan \begin{bmatrix} \left(1-\frac{t}{\tau/2}\right)\tan ^{-1}\left( {- \beta_+}/{\gamma_+} \right) \hfill \\
         +\frac{t}{\tau/2} \tan ^{-1}\left( {(-\mu_0- \beta_+)}/{\gamma_+} \right) \end{bmatrix}, & t>\tau/2,
    \end{array} \right. \label{eq:optimised_protocol}
\end{align}
where 
\begin{align}
\label{eq:beta_gamma_coefs_Kitaev}
        \beta_- &= -2\omega \cos \left( \pi/N \right), \nonumber\\
    \beta_+ &= -2\omega \cos \left((N-1) \pi/N \right),\nonumber\\
    \gamma_- &= 2|\Delta|\sin \left(\pi/N \right),\nonumber\\
    \gamma_+ &=2|\Delta|\sin \left( (N-1) \pi/N \right).
\end{align}

We note that this solution is equivalent to splitting the evolution from $\mu(0)$ to $\mu(\tau/2)=0$ and finding the optimal protocol that avoids $\beta_-$, and then evolving the system from $\mu(\tau/2)=0$ to $\mu(\tau)=$, avoiding $\beta_+$. On the other hand, if the same protocol starts on the positive side, with $\mu(0)=\mu_0'>2\omega$ and ends on the negative side, with $\mu(\tau)=-\mu_0'<-2\omega$, we have to make the changes $\beta_- \leftrightarrow \beta_+$ and $\gamma_- \leftrightarrow \gamma_+$ on the Eqs. (\ref{eq:optimised_protocol}), avoiding first $\beta_+$ and then $\beta_-$.

\section{Results}
\label{sec:results}
\subsection{Fidelity}

In this work, we investigate the dynamics of a driven chain initially prepared in its ground state, $\ket{\phi_0(0)}$. To evaluate the effectiveness of the optimized protocols, we use a linear ramp as a baseline for comparison:
\begin{equation}
    \mu(t) = \mu_0 + (\mu_f - \mu_0) t/ \tau,
\end{equation}
where $\tau$ is the total protocol duration. The time-evolved state is given by $\ket{\psi(\tau)} = \hat{U}(\tau) \ket{\phi_0(0)}$, where the unitary evolution operator is defined as
\begin{equation}
\hat{U}(\tau) = \hat{\mathcal{T}} \exp \left[ -\frac{i}{\hbar} \int_0^\tau \hat{H}(t) dt \right],
\end{equation}
and $\hat{\mathcal{T}}$ denotes the Dyson time-ordering operator. To quantify the performance of each protocol, we adopt the fidelity with the target instantaneous ground state, $\ket{\phi_0(\tau)}$, as our primary figure of merit,
\begin{equation}
    \mathcal{F} = |\langle \psi(\tau) | \phi_0(\tau) \rangle|^2.
\end{equation}
Throughout this study, we assume the initial condition $\ket{\psi(0)} = \ket{\phi_0(0)}$ and use the bulk approximation to calculate the fidelities, unless explicitly stated otherwise.

We compare the effectiveness of the two-plateau MA-STA control (Eq. (\ref{eq:optimised_protocol})) with an evolution driving the system through a linear ramp between the initial and final drives.

\begin{figure}
    \centering

    {\includegraphics[width=0.45\textwidth]{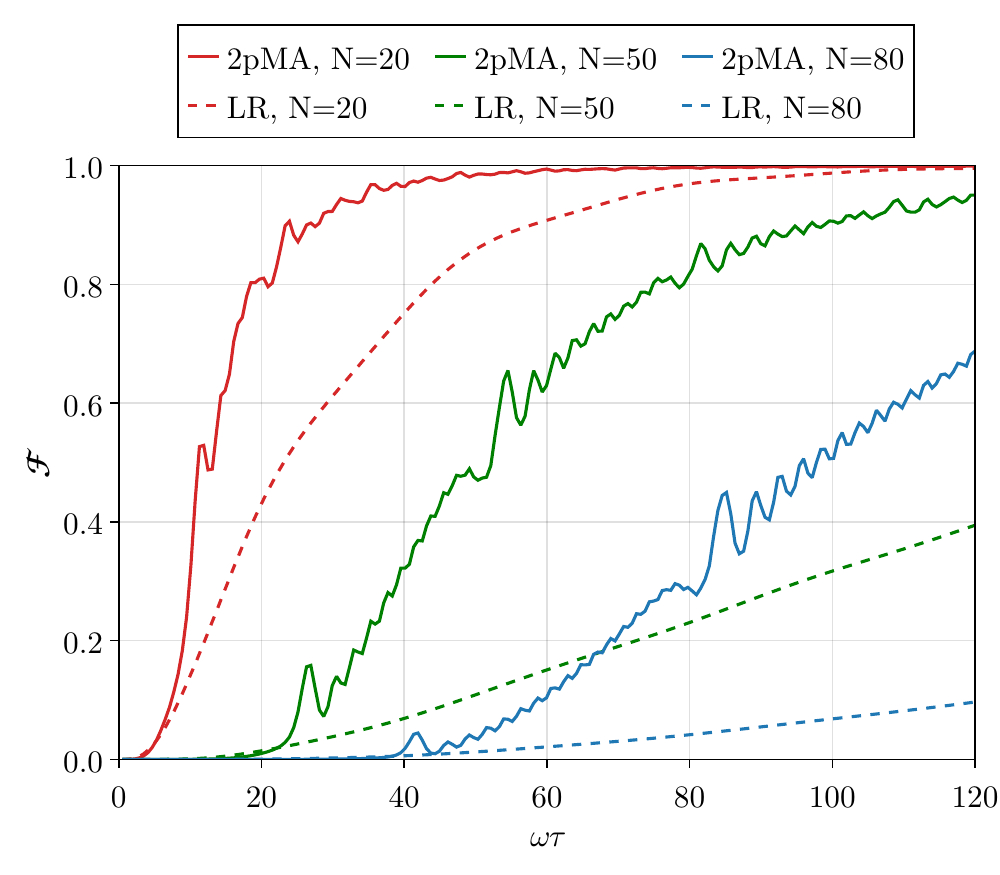}}
    
    \caption{Comparison between the two-plateaus protocol (2pMA, solid lines) and the linear ramp (LR, dashed lines) for $|\Delta| = \omega$, driven between two trivial phases and crossing two transitions with $\mu_0=-\mu_f=-3\omega$. System sizes $N = 20, 50$ and $80$ are shown in red, green and blue, respectively. The minimal action protocol demonstrates superior performance, achieving high fidelity in significantly shorter time scales compared to the linear ramp across all considered system sizes.}
    \label{fig:2nMASTA_fidelities_n_20_50_80}
\end{figure}

Fig. \ref{fig:2nMASTA_fidelities_n_20_50_80} shows the evolution up to $\omega \tau=120$ of the fidelity for Kitaev chains of lengths $N=20,50,80$ driven from $\mu_0 = -3\omega$ towards $\mu_f = 3\omega$ with constant superconducting gap $\Delta=\omega$. As discussed, it is necessary to tune the value of the gap $\Delta$ such that there are no relevant gaps other than $\Gamma_{\pi/N}$ and $\Gamma_{(N-1)\pi/N}$ on the evolution, and the effect of non-tuned values is shown on the Appendix \ref{sec:appendix_DELTA}.

For all system sizes, we observe a significant improvement of the fidelity over the linear ramp. As expected, larger systems require larger time scales to approach the adiabatic limit: while fidelities above 80\% are reached around $\omega \tau \sim 20$ for $N = 20$, chains of size $N = 80$ require at least $\omega \tau \sim 120$ to achieve 60\%.


    

\begin{figure}
    \centering

    \subfloat[]{\includegraphics[width=0.45\textwidth]{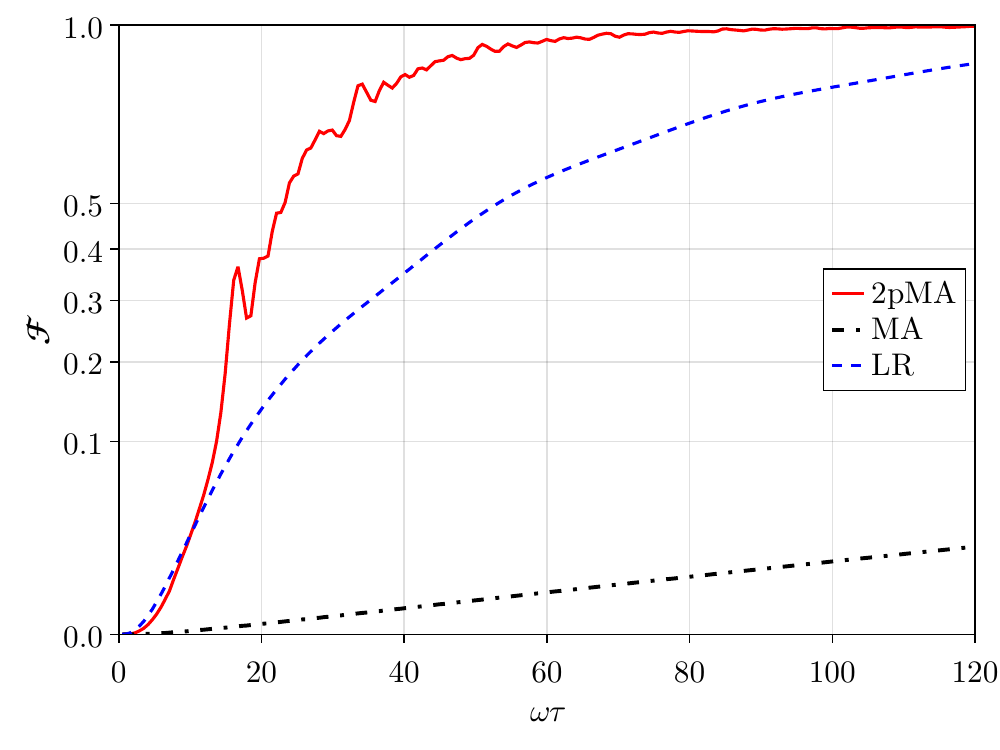}}

    \subfloat[]{\includegraphics[width=0.45\textwidth]{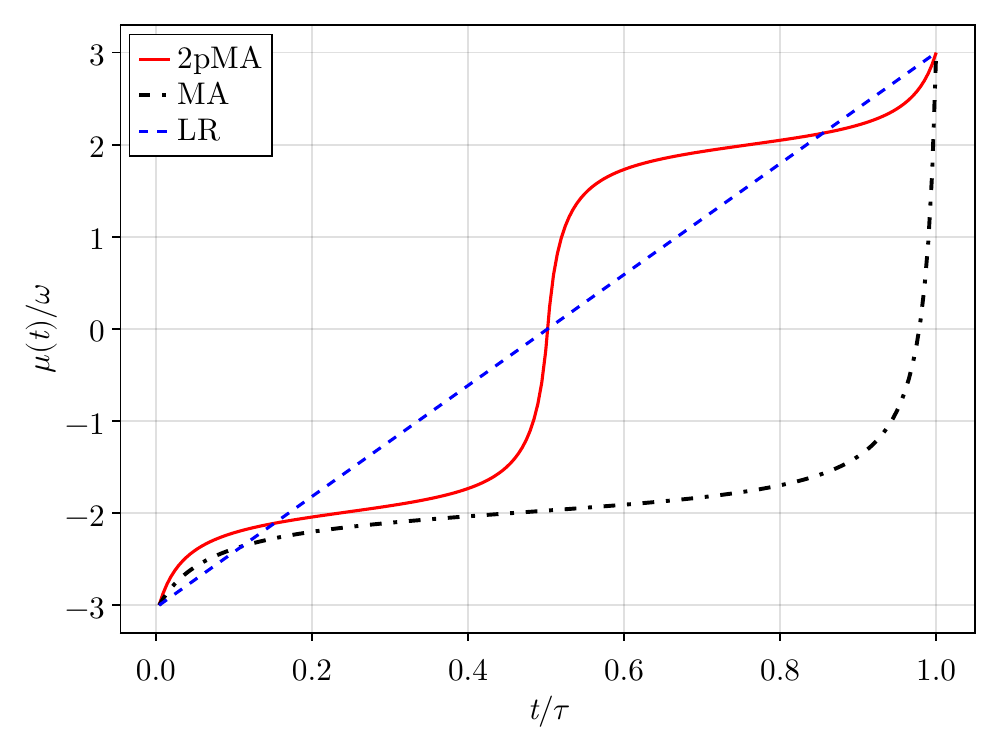}}
    
    \caption{Fidelities (a) and time-dependent control protocol $\mu(t)/\omega$ (b) for a Kitaev chain with $N = 30$ and $|\Delta| = \omega$. The quench is performed from $\mu_0 = 3\omega$ to $\mu_f = -3\omega$. Panel (a) shows that the MA-STA protocol offers a significant advantage over the standard linear ramp (LR, blue dashed lines) in achieving high-fidelity state preparation at short time scales. However, within this framework, the two-plateau optimization (2pMA, solid red lines, illustrated on the right of Fig. \ref{fig:ma_sta_prot_drawing}) is essential, as the single-plateau protocol (MA, dot-dashed black lines, left of Fig. \ref{fig:ma_sta_prot_drawing}) is insufficient and fails to converge to the target state even for longer durations. This highlights that the additional control parameters demanded by the two-step optimization are required to effectively bypass non-adiabatic excitations. Panel (b) illustrates the functional forms of the control protocol for each strategy.}
    
    \label{fig:comparison_fidelities_mu_n_30}
\end{figure}

On the other hand, if we implement the protocol while accounting for only one of the gap closures, the results are highly suboptimal even for large $\tau$, as illustrated in Fig. \ref{fig:comparison_fidelities_mu_n_30}. Although both protocols exhibit similar slopes during most of the time evolution, the overall performance depends on maintaining a low rate of change specifically when the system is near its critical points.

In scenarios where only a single phase transition is crossed and a single momentum space is involved, applying the protocol once and restricting the MA-STA minimization in the involved sector would suffice. In principle, one would expect that the simple protocol would still yield excellent results. To demonstrate this behavior, we consider a quench from the topological phase ($\mu_0=0$) to the trivial phase ($\mu_f=-3\omega$). Both the even and odd ground states, $\ket{\phi_{even/odd}}$, are evolved using the simple protocol while avoiding the most relevant gap $\Gamma_{\pi/N}$. The evolved states are given by
\begin{equation}
    \ket{\psi_{even/odd}(\tau)} = \hat{\mathcal{T}} \exp \left[ - i \int_0^\tau \hat{H}(t)dt \right]\ket{\phi_{even/odd}(0)},
\end{equation}
where $\hat{\mathcal{T}}$ is the time-ordering operator. As discussed in Section \ref{sec:kitaev_chain}, due to parity conservation, the target state for the evolved even ground state is the instantaneous ground state at the end of the evolution, $\ket{\phi_0(\tau)}$, resulting in the fidelity
\begin{align}
    \mathcal{F}_{even}(\tau)&=|\braket{\phi_0 (\tau)| \psi_{even}(\tau)}|^2,
\end{align}
Conversely, the target state for the evolved odd ground state is the first excited state, $\ket{\phi_1(\tau)}$, which yields
\begin{align}
    \mathcal{F}_{odd}(\tau)&=|\braket{\phi_1 (\tau)| \psi_{odd}(\tau)}|^2.
\end{align}

\begin{figure}
    \centering
    
    \subfloat[]{\includegraphics[width=0.45\textwidth]{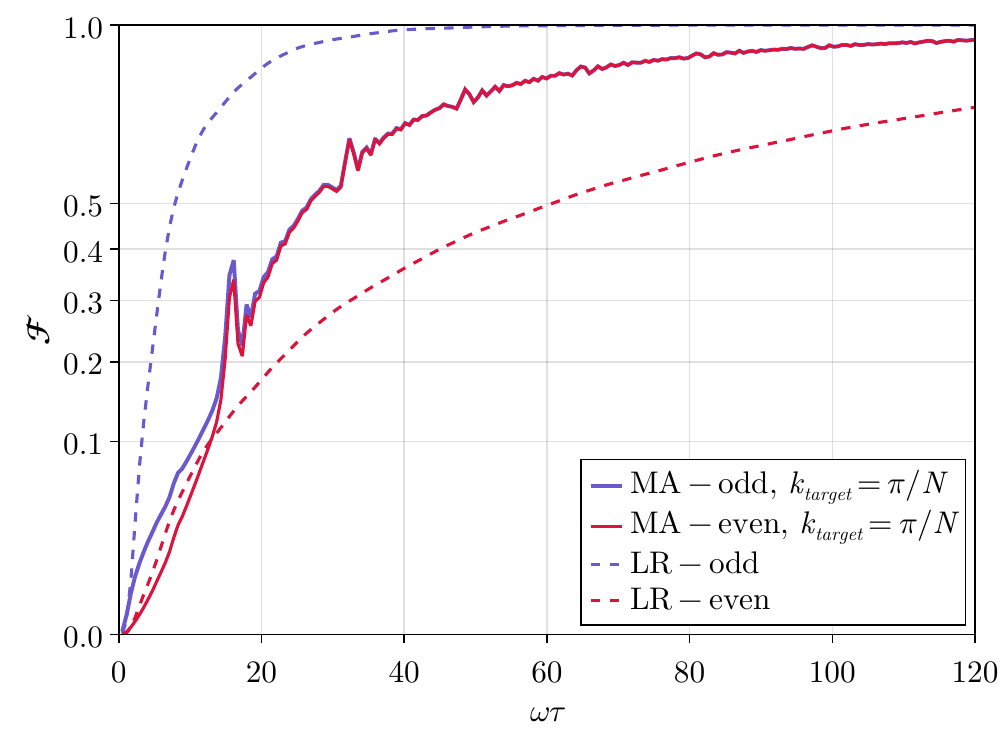}
    \label{fig:fidelity_firstmomentum_deggs_n60}}

    \subfloat[]{\includegraphics[width=0.45\textwidth]{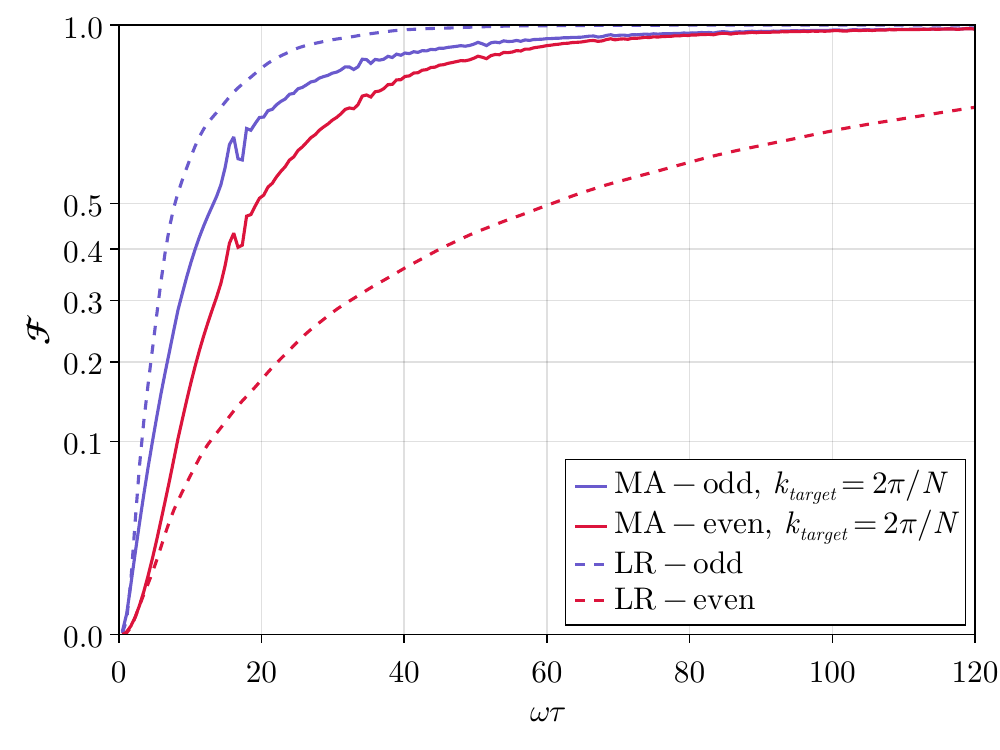}
    \label{fig:fidelity_secondmomentum_deggs_n60}}
    
    \caption{Fidelities for $|\Delta| = \omega$, $N = 60$, and a quench from $\mu_0 = 0$ to $\mu_f = -3\omega$. The plots compare the MA-STA protocol (solid lines) with the linear ramp (dashed lines) for the even (red) and odd (blue) ground states. In (a), the control is optimized for the lowest-energy momentum subspace ($k_{target}=\pi/N$), while in (b), it targets the second-lowest energy subspace ($k_{target}=2\pi/N$). Notably, the odd ground state fidelity is inherently higher due to its trivial dynamics in the odd sector. Furthermore, panel (b) reveals an unexpected improvement in the even ground state fidelity when targeting the second-lowest subspace, suggesting that optimal many-body control requires accounting for multiple energy gaps.}
    \label{fig:fidelity_deggs_n60}
    
\end{figure}

The advantage of the MA protocol over the linear ramp is again observed in Fig. \ref{fig:fidelity_firstmomentum_deggs_n60} for the even ground state, as expected. However, for the odd ground state, the linear ramp exhibits better performance. Furthermore, the fidelity achieved by the optimal protocol is nearly identical for both even and odd sectors.

\begin{figure}
    \centering

    \subfloat[]{\includegraphics[width=0.45\textwidth]{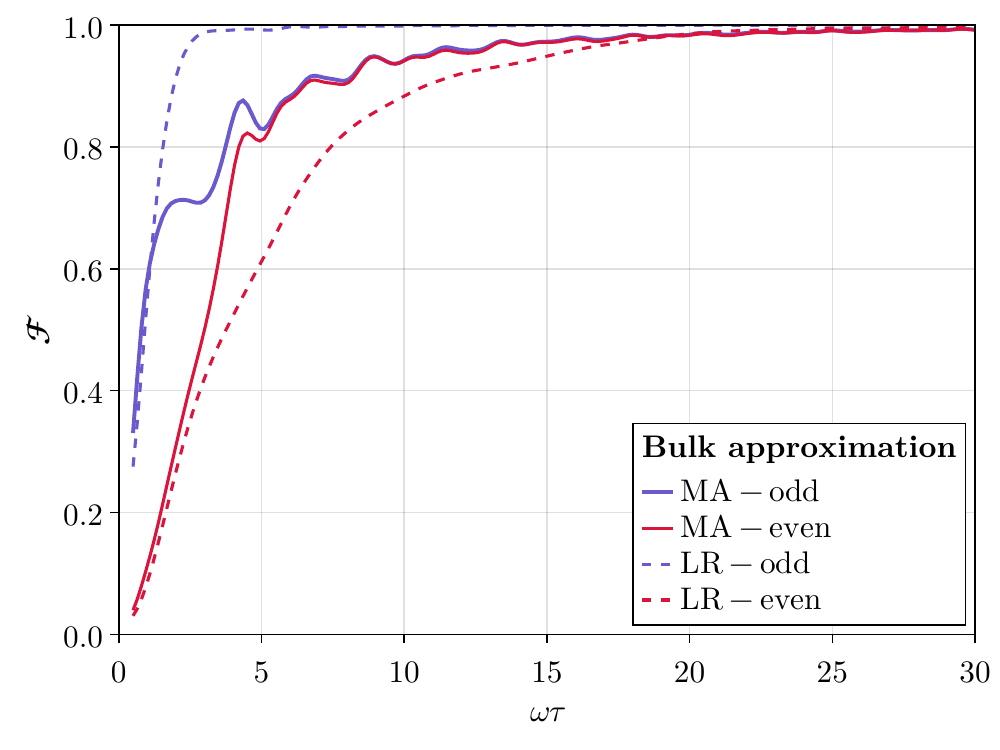}}

    \subfloat[]{\includegraphics[width=0.45\textwidth]{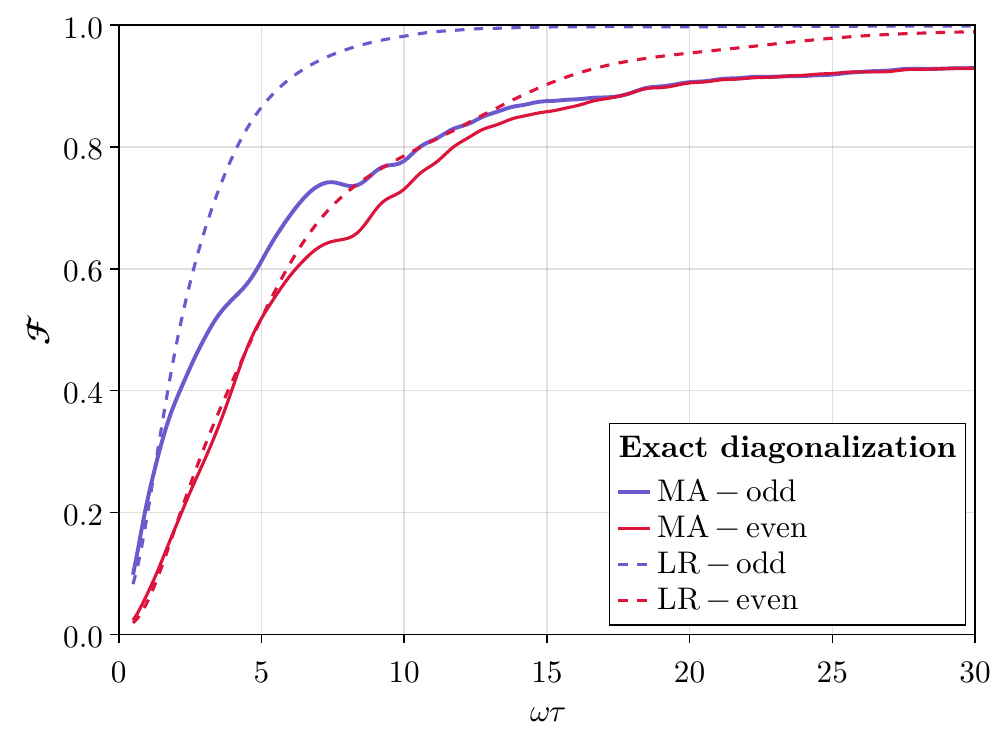}}  
        
    \caption{Fidelities for $|\Delta| = \omega$, $N = 14$, and a quench from $\mu_0 = 0$ to $\mu_f = -3\omega$. The plots compare the MA-STA protocol (solid lines) with the linear ramp (dashed lines) for the even (red) and odd (blue) ground states. In (a), the fidelity is calculated using the bulk approximation, while in (b), results are obtained via exact diagonalization. Although the bulk model predicts a clear advantage for the MA-STA protocol, the exact dynamics in (b) reveal that the linear ramp can perform better for small system sizes. These quantitative discrepancies arise because the bulk approximation neglects the discrete nature of the energy spectrum and finite-size effects, which are dominant at $N = 14$.}
    \label{fig:fidelity_comparison_n14}
\end{figure}

To ensure consistency, the same dynamics -- with $N=14$ and a total duration up to $\omega\tau=30$ -- were simulated both with and without the bulk approximation. In the latter case, the full Hamiltonian, represented by a $2^N \times 2^N$ matrix, was used to compute the time evolution via exact diagonalization in full Hilbert space, as shown in Fig. \ref{fig:fidelity_comparison_n14}. Both simulations recover the same phenomenon: the linear ramp yields higher fidelity for the odd state. The quantitative discrepancies between the two methods arise primarily because $N=14$ is not sufficiently large for the bulk approximation to be indistinguishable from the exact results. Nevertheless, the full simulation reveals that, when the MA-STA protocol is restricted to controlling only the lowest-energy momentum subspace, the linear ramp remains a superior alternative for small chains. This suggests that the trade-off of improving a single subspace at the expense of others is not beneficial in this regime, indicating that many discrete energy levels in small systems are similarly relevant to the overall performance.

As for the broader comparison between parity sectors, the similarity in the behavior of the even and odd fidelity under the MA-STA protocol is not accidental. As demonstrated in Appendix \ref{sec:appendix_A}, the fidelities of the even and odd ground states differ only in the lowest-energy momentum subspace. Since the odd ground state and the first excited state have their critical momentum components in the odd sector -- where the evolution operator $\hat{U}_{\pi/N,odd}$ contributes only a global phase --, this specific subspace requires no active control and maintains maximum fidelity for all $\tau$. As shown in Fig. \ref{fig:fidelity_firstmomentum_deggs_n60},  once the relevant even-sector subspaces are controlled, the two curves converge.

Since the lowest-energy subspace for the odd ground state requires no active control, an advantage over the linear ramp could be recovered by considering the second-lowest energy gap as the relevant scale for the dynamics. However, as shown in Fig. \ref{fig:fidelity_secondmomentum_deggs_n60}, while the fidelity of the minimal action protocol does increase, it underperforms the linear ramp for odd ground-state. A resoning for that is that, for this state, multiple subspaces are similarly relevant; thus, improving the fidelity of a single subspace at the expense of others proves to be highly unfavorable.

\begin{figure*}
    \centering

    \subfloat[]{\includegraphics[width=0.4\textwidth]{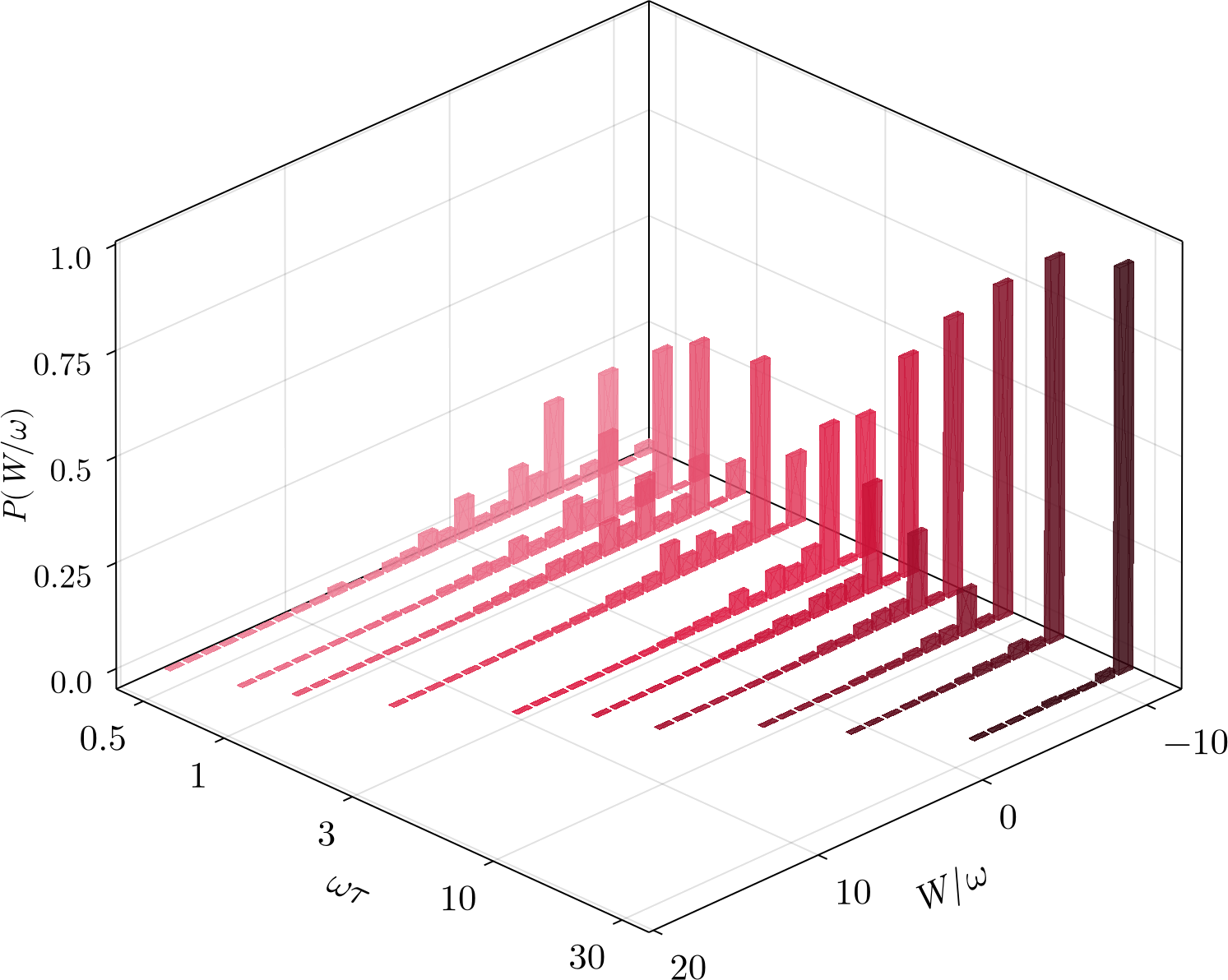}
    \label{fig:work_distros_even}}
    \subfloat[]{\includegraphics[width=0.4\textwidth]{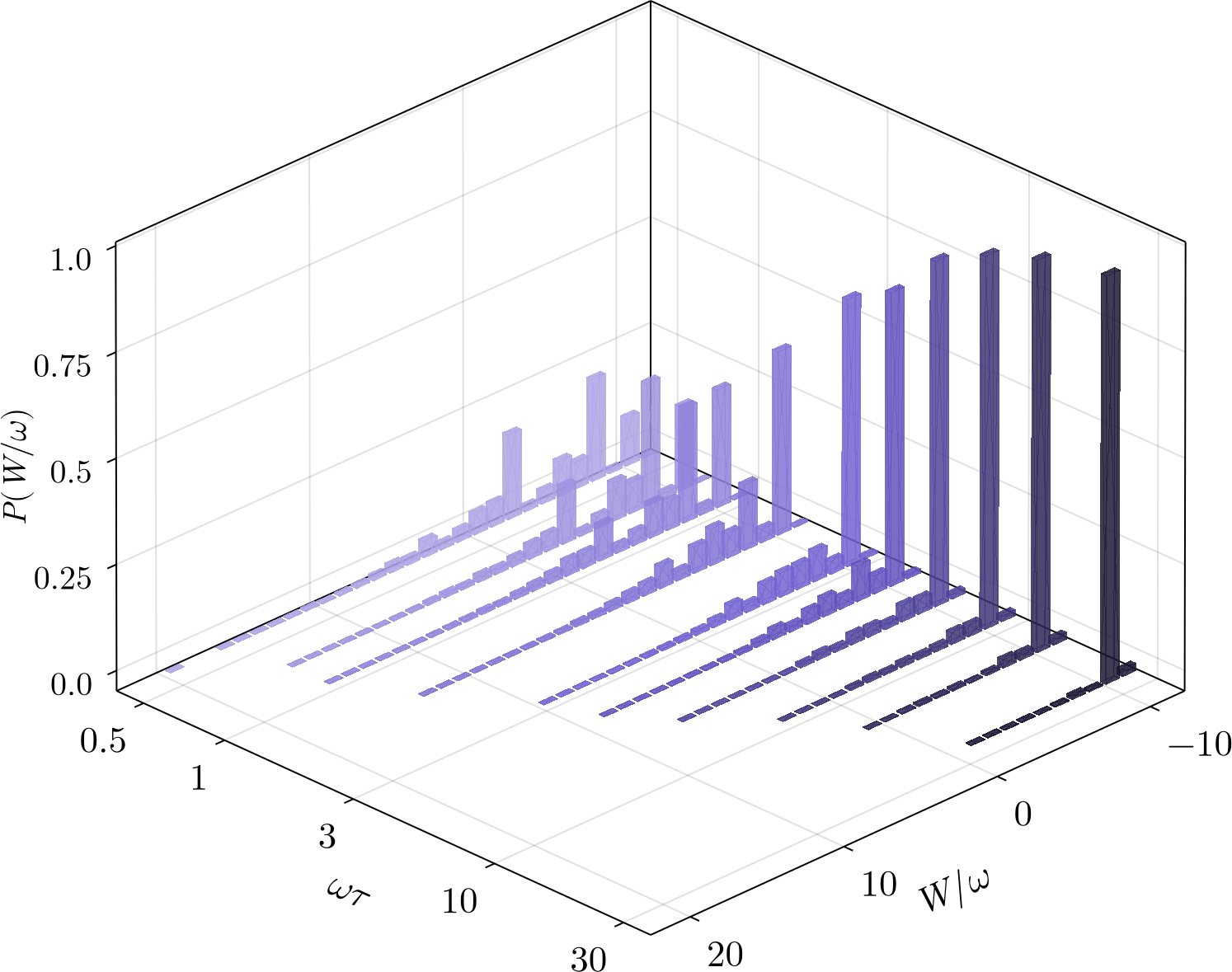}
    \label{fig:work_distros_odd}}
   
    \subfloat[]{\includegraphics[width=0.33\textwidth]{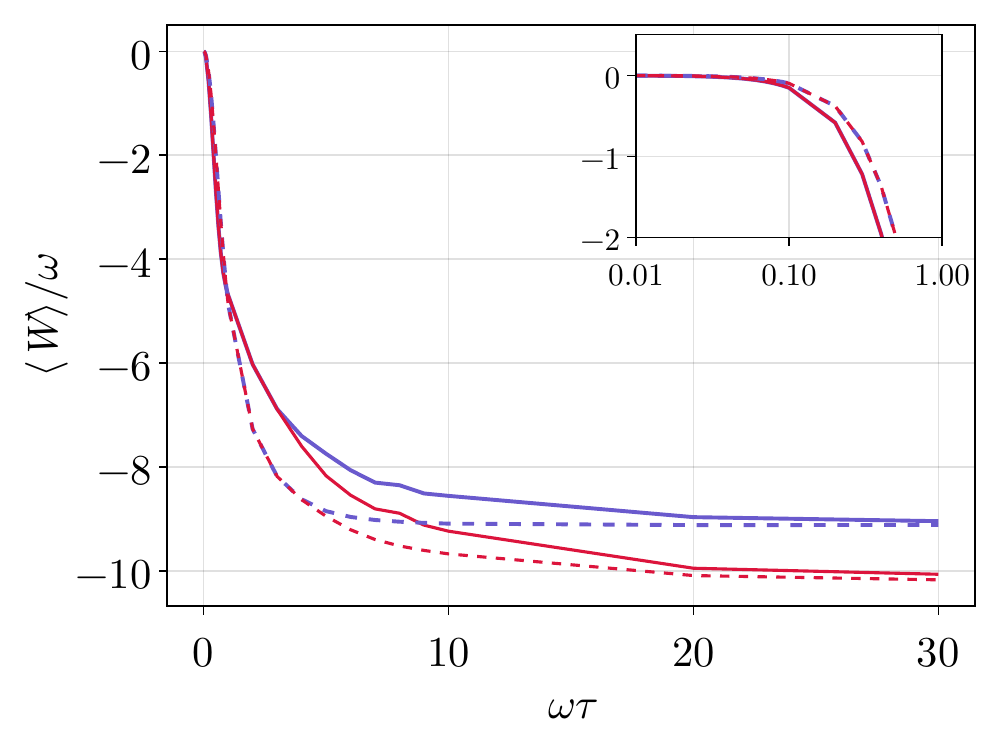}
    \label{fig:work_distros_mean}}
    \subfloat[]{\includegraphics[width=0.33\textwidth]{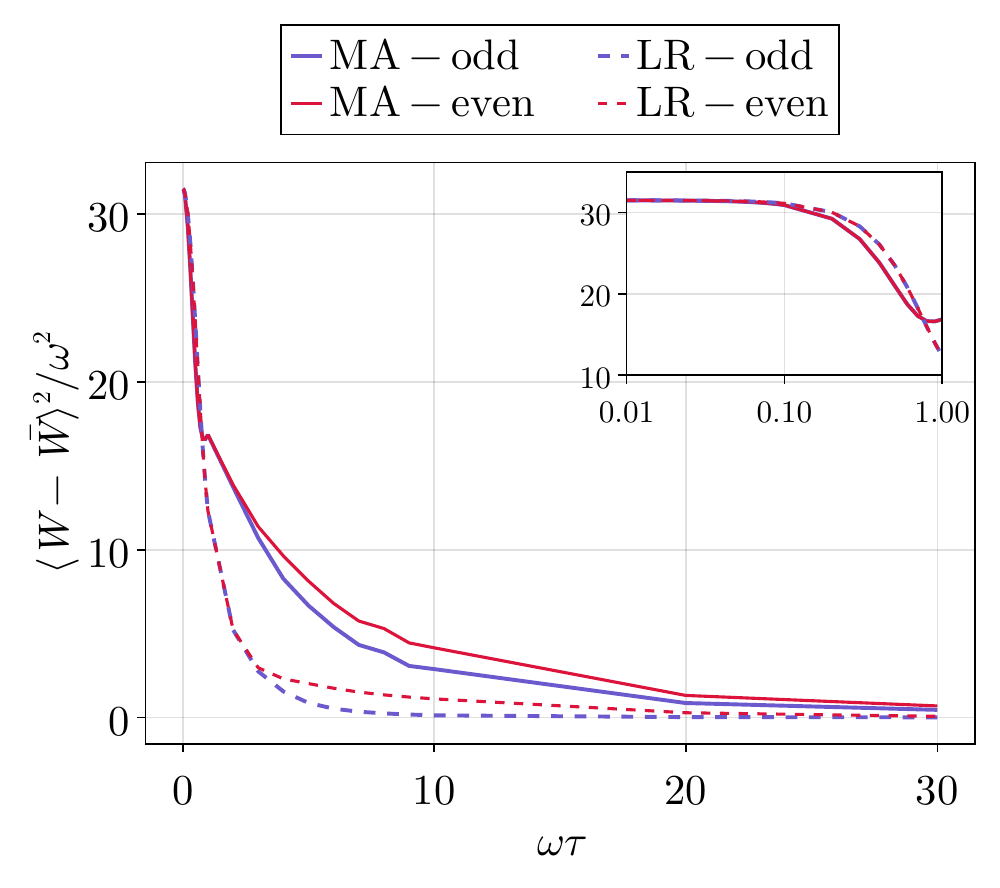}
    \label{fig:work_distros_variance}}
    \subfloat[]{\includegraphics[width=0.33\textwidth]{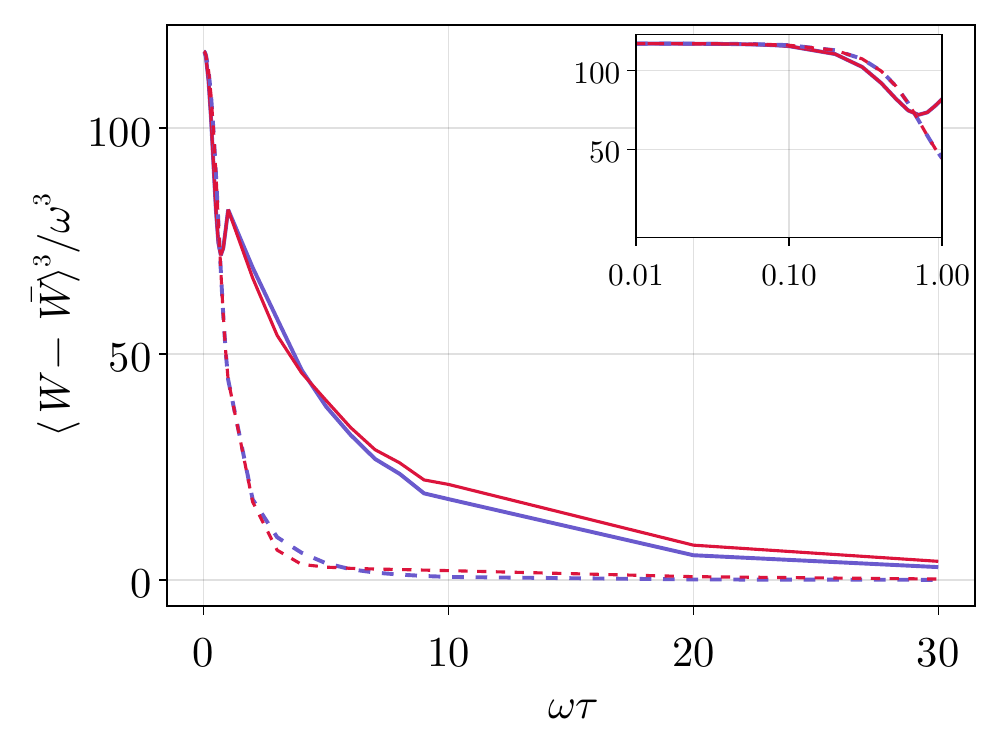}
    \label{fig:work_distros_skewness}}
    
    \caption{Work probability distribution $P(W)$ for the MA-STA protocol in (a) and (b), and distribution moments ((c), (d) and (e)) as a function of the quench duration $\omega\tau$. Results are obtained via exact diagonalization of a Kitaev spin chain with $N = 14$ sites and a quench from $\mu_0 = 0$ to $\mu_f = -3\omega$. Panels (a) and (b) show the evolution of the work distribution starting from the even and odd ground states, respectively. Panels (c), (d) and (e) show, respectively, the first, second and third moments of the distributions using the MA protocol (solid lines) and the linear ramp (dashed lines) for both the even (red) and odd (blue) ground states. Insets in (c)-(e) highlight the sudden quench regime ($\omega\tau \ll 1$) on a logarithmic time scale, revealing work fluctuations associated with transitions to excited states. As $\tau$ increases, the probability weight concentrates into adiabatic peaks: the system converges to the final ground state in the even sector (a), and to the first excited state in the odd sector (b). The energy shift between these peaks corresponds to the topological gap energy of the model.}
    \label{fig:work_distros}
\end{figure*}

Interestingly, targeting the second-lowest momentum subspace yielded significantly better results for the even ground state, reaching a fidelity of 50\% at approximately $\omega\tau \approx 20$. This result can be understood by noting that the total fidelity is a product of contributions from all momentum subspaces, as shown in Appendix \ref{sec:appendix_A}. Optimizing the protocol for the lowest-energy subspace ensures near-adiabatic evolution in that sector, but does so at the cost of accelerating the drive through the remaining ones. If the second-lowest subspace is itself close to a gap closure, this acceleration may render it the dominant source of non-adiabatic excitations. By targeting the second-lowest subspace instead, the protocol distributes the control effort more evenly across the two most critical sectors, yielding a better overall fidelity. This suggests that the optimal control strategy should not rely solely on the minimum energy gap, but instead account for a balance between the most relevant momentum sectors, perhaps through a weighted adiabatic action incorporating multiple energy scales simultaneously.

\subsection{Work distribution and fluctuations}

Beyond the fidelity, we characterize the energy exchange and fluctuations during the process through the work distribution, obtained via the standard two-point measurement scheme \cite{Talkner_2007}. In this approach, work is associated with a stochastic variable whose values correspond to the transition energies between states of the initial and final Hamiltonians, $W=\varepsilon^\tau_m-\varepsilon^0_n$. Each transition occur with a probability
\begin{equation}
    P(W)=\sum_{n,m}p_n^0p_{m|n}^\tau \delta \left[ W-(\varepsilon^\tau_m-\varepsilon^0_n) \right],
\end{equation}
where $p_n^0$ is the initial-state occupation probability of the eigenstate $\ket{n^0}$, with energy $\varepsilon_n^0$, and $p_{m|n}^\tau=|\bra{m^\tau} \hat{U}(\tau) \ket{n^0}|^2$ represents the transition probability from an initial state $\ket{n^0}$ to the final eigenstate $\ket{m^\tau}$.

In the adiabatic limit, the transition probability is $p_{m|n}^\tau=\delta_{m,n}$, whereas in the sudden quench regime, it reduces to the overlap $p_{m|n}^\tau=|\braket{m^\tau | n^0}|^2$. Since the system initiates in the ground state, the initial-state occupation probability is $p_n^0=\delta_{n,0}$ and the work distribution simplifies to 
\begin{equation}
P(W)=\sum_{m}|\bra{m^\tau} \hat{U}(\tau) \ket{\phi_0(0)}|^2 \delta \left[ W-(\varepsilon^\tau_m-\varepsilon^0_0) \right].
\end{equation}

The statistics of the work distribution defined above, in particular the second moment, allows for an analysis of the work fluctuations.  The $k$-th moments of the distribution are defined as:
\begin{equation}
    \bar{\mathcal{W}}^k = \langle (W-\bar{W})^k\rangle=\sum_i P(W_i)(W_i - \bar{W})^k.
\end{equation}

We analyze the evolution of the work probability distribution $P(W)$ as a function of the quench duration $\omega\tau$ in Fig.~ \ref{fig:work_distros}. As expected, for short protocols ($\omega\tau \ll 1$), the distribution is broad, with significant weight distributed across higher energy transitions. This behavior is typical of a sudden quench \cite{Zawadzki_2023}, where the fast drive populates a wide range of excited states, leading to irreversible work production. By increasing the protocol duration $\tau$, the off-diagonal transitions are progressively suppressed.
In the even parity sector (Fig. \ref{fig:work_distros_even}), the distribution localizes around the work value corresponding to the energy difference between the final and initial ground states $W = E_0^\tau - E_0^0$. The peak in the odd parity sector (Fig. \ref{fig:work_distros_odd}) is  shifted by the topological gap $\Gamma_{\pi/N}$ of the Kitaev chain. This reflects the fact that the initial odd ground state evolves toward the first excited state of the final Hamiltonian within the parity-protected manifold.

In an adiabatic dynamics, one would expect a sharp work distribution peaked at an average work value close to the transition between the initial and the target eigenstates. Therefore, we can analyze the moments of the work distribution under the MA-STA protocol and linear ramp protocols.
The moments $k=1$ (mean, non-centered), $k=2$ (variance) and $k=3$ (skewness) are shown, respectively on Figs. \ref{fig:work_distros} (c), (d) and (e) as a function of the quench duration. The short-time dynamics, highlighted in the corresponding insets on a logarithmic scale, reveal that for very short protocols ($\omega \tau \ll 1$), all moments for both the MA-STA and the linear ramp reflect an asymmetric and broad distribution centered near zero. In particular, as the protocol duration increases, the average work $\langle W \rangle$ rapidly converges to the adiabatic energy differences $\varepsilon_0^\tau - \varepsilon_0^0$ for the even ground state and $\varepsilon_1^\tau - \varepsilon_0^0$ for the odd.

By comparing work fluctuations and the skewness of the distributions obtained from the MA-STA and the LR protocol, we can clearly see that the latter outperforms. This behavior is consistent with the results observed for the fidelity (\ref{fig:fidelity_comparison_n14}) and can be attributed to finite-size effects in small chains, for which the bulk approximation may need some adjustments. Our results for relatively large chains suggest that it is necessary to conduct a finite-size scaling analysis of the work statistics. 


\section{Conclusions}
\label{sec:conclusions}

In this work, we have investigated the performance of the Minimal Action Shortcut to Adiabaticity (MA-STA) in the Kitaev chain, evaluating its capacity to drive the system between the trivial and topological phases at finite time. Our results suggest that the performance of the shortcut protocol is intrinsically tied to specific momentum-dependent gap closures.
In particular, we demonstrated that for some values of the pairing interaction, with $|\Delta| \neq \omega$, the relevant gap can be found in a different momentum sector other than $k=0$ and $k=\pi$.  In the weak-pairing limit ($|\Delta| \ll \omega$), the energy gap remains small for a wide range of the driving parameter $\mu(t)$, leading to an unexpected minimal gap associated with specific momentum sectors ($k_{S_3}$). In order to adapt the MA-STA to situations in which this persistent proximity of excited states occur in momentum sectors, one has to go beyond the single-gap estimation.

 For the Kitaev chain, we showed that to be the case of a transition in which the entire topological phase is crossed to connect two trivial phases for $\mu < -2\omega$ and $\mu>2\omega$, a two-plateau strategy is essential. By employing a bulk-spectrum approximation, we observed that the odd-parity ground state is inherently easier to control, as its lowest-energy momentum subspace requires no control at all. Consequently, the most efficient control strategy involves targeting the even momentum sector with the smallest gap (the second-lowest momentum mode). Remarkably, optimizing for this specific subspace also yields more robust results for the even ground state. This suggests that a truly effective shortcut must go beyond the absolute minimal gap and instead account for a broader range of momentum subspaces. A future promising direction in refining the MA-STA framework is to define the action in terms of weighted contributions of gaps between low-lying states from different sectors.

\section{Acknowledgments}
The authors thank G. Diniz, A. Kiely, S. Campbell, M.F. Cavalcante and D. Rosa for providing insightful comments on early versions of the present manuscript.
R.B.S. acknowledges the São Paulo Research Foundation (FAPESP) for financial support (Grant 2025/06385-9). K.Z. acknowledges CNPq Grant
No.305665/2025-1.


\appendix
\begin{appendices}
\section{Influence of non-tuned $\Delta$ on the effectiveness of the MA-STA}
\label{sec:appendix_DELTA}

\begin{figure}
    \centering

    \subfloat[]{\includegraphics[width=0.45\textwidth]{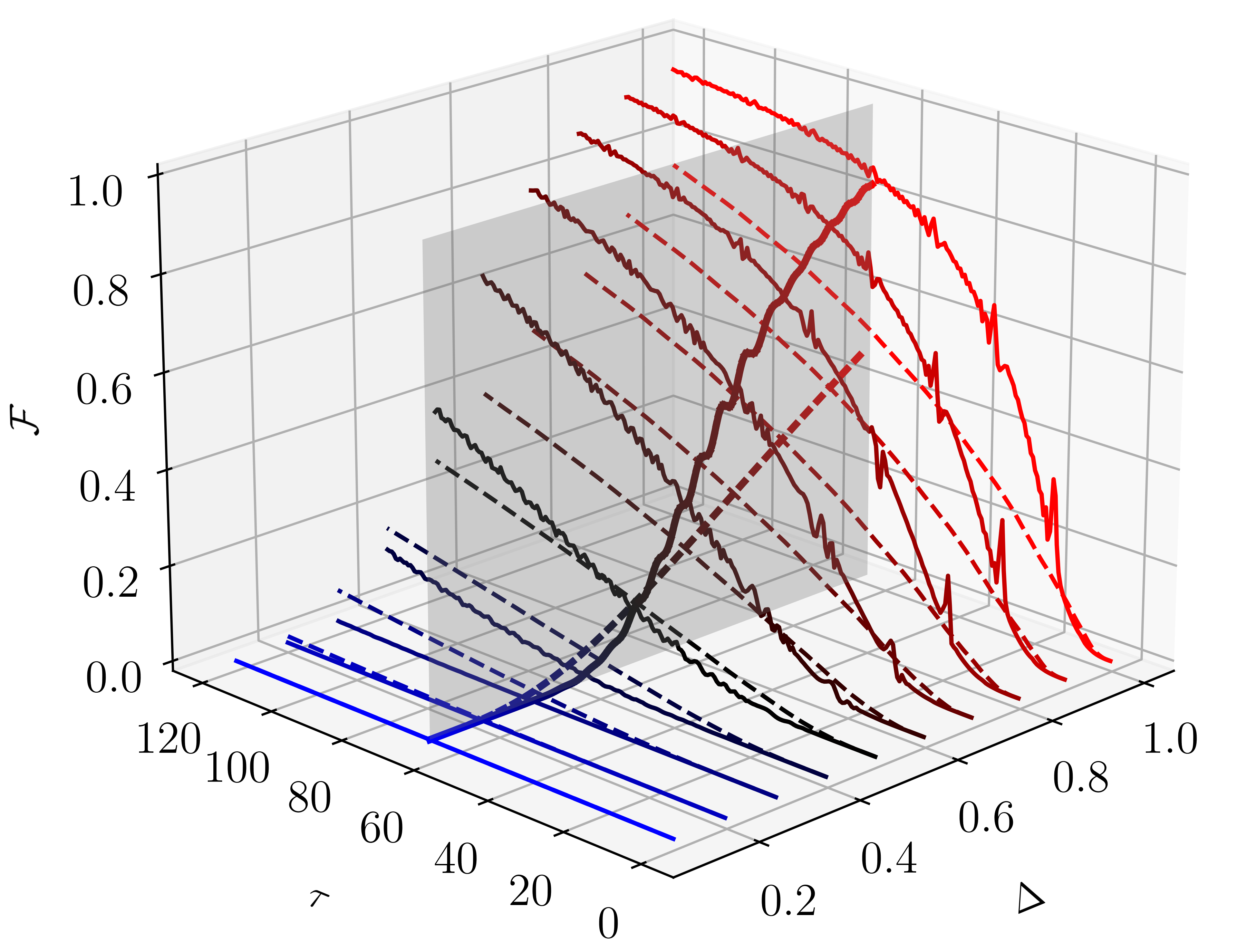}}

    \subfloat[]{\includegraphics[width=0.45\textwidth]{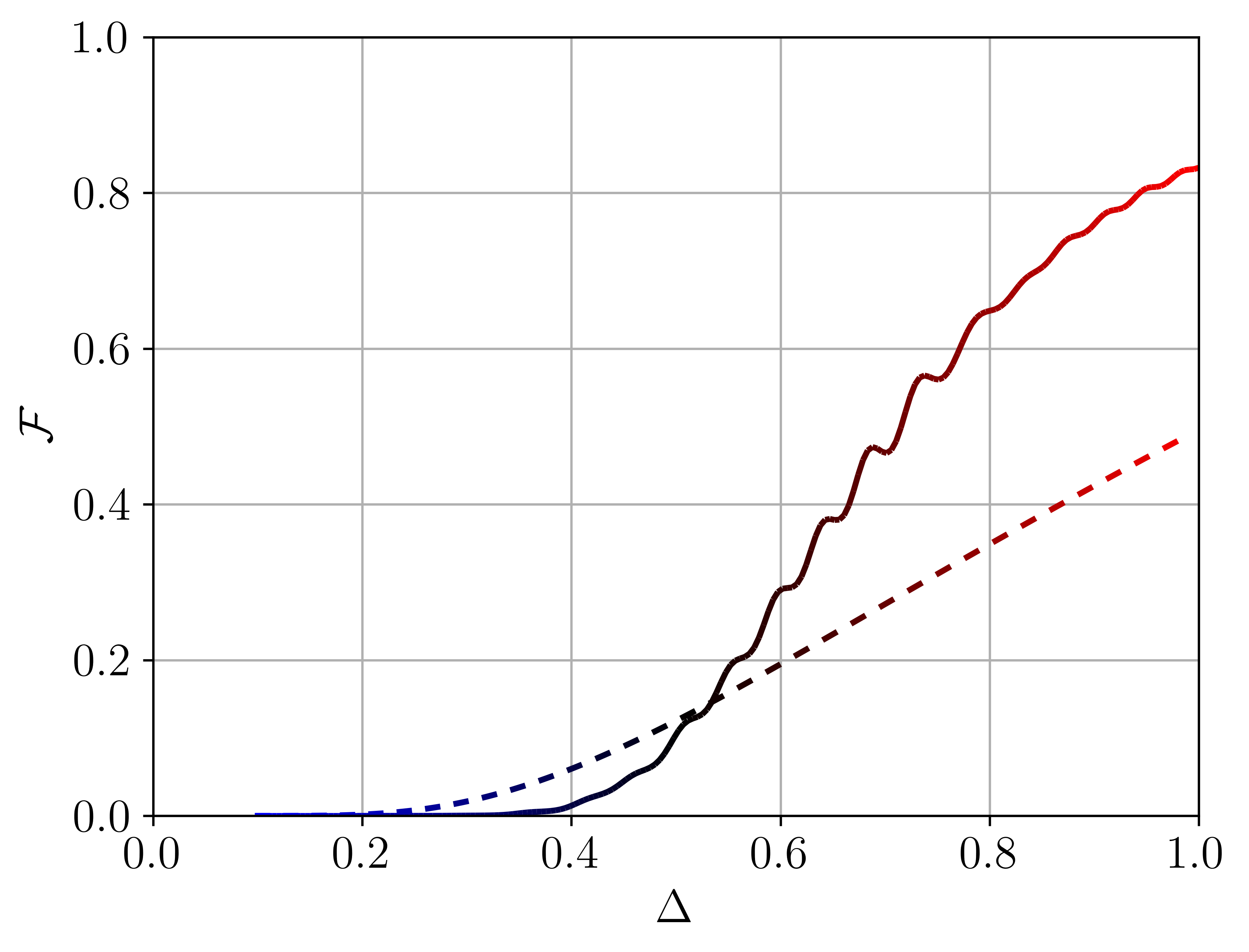}}
    
    \caption{Fidelity for various values of $\Delta$ with $\mu_0=-3\omega$, $\mu_f=0$ and $N=60$. The total time goes up to $\omega\tau=120$ (a), and at $\omega\tau=60$ (gray plane) the fidelities are plotted for continuously varying $\Delta$ (b). Results' colors change from blue for $\Delta=0.1\omega$, to black for $\Delta=0.5\omega$ and to red for $\Delta=\omega$, with solid lines for the MA-STA protocol and dashed for the simple linear ramp.}
    
    \label{fig:fidel_vs_delta}

\end{figure}

The superconducting gap $\Delta$ is highly related to the minimal energy gap of each subspace through $\gamma=2|\Delta|\sin k$, as defined on Equation \ref{eq:generalization_adaction}, so it is expected that smaller values worsen the overall performance of both protocols. However, we have shown that for sufficiently small values of $\Delta$, non-expected minimal gaps appear close to the chemical potential $\mu(t)$ at the topological phase, which means the MA-STA is more sensitive to the performance worsening than the linear ramp, thus making its usage disadvantageous. This regime can be seen on Fig. \ref{fig:fidel_vs_delta}, where the curves are blue and the linear ramp's dashed lines are above the minimal action's solid ones. This difference in performances becomes more significant as $N$ decreases and the total time evolution increases, and for a fixed time evolution $\omega \tau=60$ the MA-STA outperforms the linear ramp around $\Delta=0.5\omega$, and continuously gets better as the gap increases.

\section{Odd parity ground state of the Kitaev Chain in the momentum representation}
\label{sec:appendix_A}

In the topological phase, starting from the even ground state $\ket{G_{even}}$, we can construct the odd ground state by applying the non-local creation operator $\hat{f^\dagger}=(a_1^\dagger - ib_N^\dagger)/2$ in the Majorana representation, which is a parity operator that connects the even and odd sectors of the Hilbert space,
\begin{equation}
    \ket{G_{odd}}=\hat{f}^\dagger \ket{G_{even}}.
\end{equation}

Considering $\ket{k^\pm_t}$ the eigenstates of the subspace of momentum $k$, where $\pm$ denotes respectively the excited and ground states on time $t$, in the momentum basis, this state can be written as
\begin{equation}
    \ket{G_{even}(0)}=\otimes_{k=0}^\pi \ket{k_0^-},
\end{equation}
and in the Majorana representation ($\hat{a}_1^\dagger=(\hat{c}_1^\dagger+\hat{c}_1)$ and $\hat{b}_{N}^\dagger=i(\hat{c}_N^\dagger-\hat{c}_N)$), we have
\begin{align}
    \hat{f}^\dagger &= \frac{1}{2}(\hat{a}_1^\dagger-i \hat{b}_{N}^\dagger)\\
    &=\frac{1}{2}(\hat{c}_1^\dagger+\hat{c}_1+\hat{c}_N^\dagger-\hat{c}_N)\\
    &=\frac{1}{2}\sum_k \frac{1}{\sqrt{N}}\left( e^{-ik}\hat{c}_k^\dagger + e^{ik} \hat{c}_k + e^{-ikN} \hat{c}_k^\dagger - e^{ikN}\hat{c}_k \right),
\end{align}
which, by taking $k<0$ out of the summation, yields
\begin{align}
    \begin{split}
        \hat{f}^\dagger=\frac{1}{2}\sum_{k \geq 0} \frac{1}{\sqrt{N}} \bigl( e^{-ik}\hat{c}_k^\dagger + e^{ik} \hat{c}_k + e^{-ikN} \hat{c}_k^\dagger - e^{ikN} \hat{c}_k \\
    + e^{ik}\hat{c}_{-k}^\dagger + e^{-ik} \hat{c}_{-k} + e^{ikN} \hat{c}_{-k}^\dagger - e^{-ikN}\hat{c}_{-k} \bigr),
    \end{split}
\end{align}
so we can define
\begin{equation}
    \hat{f}^\dagger = \sum_{k\geq 0}\hat{f}^\dagger_k,
\end{equation}
where $\hat{f}^\dagger_k$ acts only on the subspace of momentum $k$. With this, we can write the odd ground state as
\begin{equation}
    \ket{G_{odd}(0)}=\hat{f}^\dagger \ket{G_{even}(0)}=\left(\sum_{k=0}^\pi \hat{f}_k^\dagger \ket{k_0^-} \otimes_{\substack{k'=0\\ k'\neq k}}^\pi \ket{k'^-_0} \right),
\end{equation}
and if the dynamics is such that $\mu_0=0$ and $\mu_f=-3\omega$, when we cross the phase transition the gap of momentum subspace $k=0$ closes, so this is also the subspace with lowest energy. Therefore, on $t=\tau$, at the end of the dynamics, at the trivial phase, the first true excited state, which has odd parity, can be calculated by applying the quasiparticle creation operator on the $k=0$ subspace. This operator is defined as
\begin{equation}
    \hat{\beta}_0=u_k \hat{c}_k + v_k \hat{c}_{-k}^\dagger,
\end{equation}
where \cite{Alicea_2012}
\begin{align}
    u_k &= \frac{\tilde{\Delta}_k}{|\tilde{\Delta}_k|} \sqrt{\frac{E_{bulk}+\varepsilon_k}{E_{bulk}}}, \\
    v_k &= \left( \frac{E_{bulk}-\varepsilon_k}{\tilde{\Delta}_k} \right) u_k,
\end{align}
with
\begin{align}
    E_{bulk}&=\sqrt{\varepsilon_k^2 + |\tilde{\Delta}_k|^2},\\
    \varepsilon_k &= -2\omega \cos k - \mu,\\
    \tilde{\Delta}_k &= -2i |\Delta|\sin k.
\end{align}

With this, we can calculate the first excited state,
\begin{equation}
    \ket{\phi_1(\tau)}=\hat{\beta}_0^\dagger \ket{\phi_0(\tau)}=(\hat{\beta}_0^\dagger\ket{0_\tau^-})\otimes_{k>0}^\pi \ket{k_0^-},
\end{equation}
and compute the evolved odd state,
\begin{equation}\begin{split}
    \psi_{odd}(\tau) &= \hat{U}(\tau) \ket{G_{odd}(0)}\\
    &=\sum_{k\geq0}^\pi  \biggl( \left( \hat{U}_k (\tau) \hat{f}_k^\dagger \ket{k_0^-} \right) \cdot \prod_{\substack{k'\geq0\\ k'\neq k}}^\pi \left( \hat{U}_{k'}(\tau) \ket{k'^-_0} \right) \biggr).
\end{split}\end{equation}

At this stage, it is necessary to express all operators and states in a common basis to ensure a consistent representation. In a $k$ momentum subspace, the states have the form $\ket{site\ +k, \; site\ -k}$, so this subspace is fully spanned by $\left\{\ket{00},\ket{11},\ket{01},\ket{10} \right\}$. However, the Kitaev chain's Hamiltonian, in the Bogoliubov de-Gennes representation, $\hat{H}_{BdG}$, which is being used directly to calculate the final fidelity, is quadratic in fermions, so it only acts on the subspace $\left\{\ket{00},\ket{11}\right\}$, the sector of the hamiltonian with even parity. On the other hand, the operator $\hat{f}^\dagger$ is not quadratic in fermions -- in fact, it acts on all basis' space and changes the parity of a state. So, to compute the fidelity, we have to consider the Hamiltonian in the total $4\times4$ space:
\begin{equation}
\begin{split}
    &\hat{H}_{BdG}^{full} = \hat{H}_{BdG}^{even} \oplus \hat{H}_{BdG}^{odd}\\
    &= -\begin{pmatrix}
        \mu + 2 \omega \cos k & -2i\Delta^* \sin k & 0 & 0\\
        2i \Delta \sin k & -\mu - 2 \omega \cos k & 0 & 0\\
        0 & 0 & {\mu + 2 \omega \cos k} & 0\\
        0 & 0 & 0 & \mu + 2 \omega \cos k
    \end{pmatrix},
\end{split}
\end{equation}
and since it is block diagonal, the total evolution operator is also block-diagonal
\begin{align}
    \hat{U}_{k,full} = \begin{pmatrix}
        \hat{U}_{k,even} & 0 \\
        0 & \hat{U}_{k,odd}
    \end{pmatrix}.
\end{align}

We can also write the matrix form of $\hat{f}^\dagger_k$ in the $\left\{\ket{00},\ket{11},\ket{01},\ket{10} \right\}$ basis by applying the operator on these states and using the result as columns,
\begin{equation}
    \hat{f}^\dagger_k = \frac{1}{2\sqrt{N}}\begin{pmatrix}
        0 & 0 & e^{-ik}-e^{-ikN} & e^{ik}-e^{ikN}\\
        0 & 0 & e^{-ik}+ e^{-ikN} & -e^{ik} - e^{ikN}\\
        e^{ik} + e^{ikN} & e^{ik} - e^{ikN} & 0 & 0\\
        e^{-ik}+e^{-ikN} & -e^{-ik} + e^{-ikN} & 0 & 0
    \end{pmatrix},
\end{equation}
and the same for the $\hat{\beta}_k$ operator,
\begin{equation}
    \hat{\beta}_k=\begin{pmatrix}
        0 & 0 & 0 & u_k\\
        0 & 0 & 0 & -v_k\\
        v_k & u_k & 0 & 0\\
        0 & 0 & 0 & 0
    \end{pmatrix},
\end{equation}
which is also an operator that takes the $k$'th momentum subspace to the odd sector of the Hamiltonian. Putting this all together, the final fidelity is
\begin{align}
    \begin{split}
        \mathcal{F}_{odd} &= \Bigg| \sum_{k\geq0}^\pi \bra{\phi_1(\tau)} \biggl( \left( \hat{U}_k (\tau) \hat{f}_k^\dagger \ket{k_0^-} \right) \cdot \prod_{\substack{k'\geq0\\ k'\neq k}}^\pi \left( \hat{U}_{k'}(\tau) \ket{k'^-_0} \right) \biggr) \Bigg|^2\\
        &=\Bigg| \left( \bra{0_\tau^-}\hat{\beta}_0\hat{U}_0(\tau) \hat{f}_0^\dagger \ket{0_0^-} \right) \cdot \prod_{k'>0}^\pi \left( \bra{k'^-_\tau} \hat{U}_{k'}(\tau) \ket{k'^-_0} \right) +\\
        &\quad\sum_{k>0}^\pi \biggl( \left(\bra{k_\tau^-} \hat{U}_k (\tau) \hat{f}_k^\dagger \ket{k_0^-} \right) \cdot \prod_{\substack{k'\geq0\\ k'\neq k}}^\pi \left( \bra{k'^-_\tau} \hat{U}_{k'}(\tau) \ket{k'^-_0} \right) \biggr) \Bigg|^2,
\end{split}\end{align}
but since $\hat{f}_k^\dagger$ maps the state to the odd sector, its overlap with the momentum components of the final first excited state vanishes for all $k \neq 0$. The only non-vanishing contribution comes from the $k=0$ subspace, which also resides in the odd sector due to the action of $\hat{\beta}^\dagger_0$. Furthermore, since $\hat{H}_{BdG}^{odd}$ remains diagonal throughout the dynamics, the evolution in this sector merely contributes a phase factor that does not affect the overall fidelity magnitude. Consequently, the final fidelity for the odd ground state simplifies to
\begin{equation}
    \mathcal{F}_{odd}= \left| \left( \bra{0_\tau^-}\hat{\beta}_0\hat{U}_{0,odd}(\tau) \hat{f}_{0}^\dagger \ket{0_0^-} \right) \cdot \prod_{k'>0}^\pi \left( \bra{k'^-_\tau} \hat{U}_{k',even}(\tau) \ket{k'^-_0} \right) \right|^2,
\end{equation}
whereas the fidelity of the even ground state is
\begin{equation}
    \mathcal{F}_{even}= \left| \prod_{k=0}^\pi \left( \bra{k^-_\tau} \hat{U}_{k,even}(\tau) \ket{k^-_0} \right) \right|^2,
\end{equation}
so they differ only on the lowest-energy momentum subspace.

\end{appendices}

\bibliography{references}

\end{document}